\documentclass[aps,prxquantum,twocolumn,notitlepage,superscriptaddress]{revtex4-2}
\usepackage{amsmath,amssymb}
\usepackage{bm}
\usepackage{natbib}
\usepackage{mathtools}
\usepackage{braket}
\usepackage[hidelinks]{hyperref}
\usepackage{graphicx}
\usepackage[dvipsnames]{xcolor}
\usepackage{color}
\usepackage[caption=false]{subfig}
\usepackage{blkarray}
\usepackage{hhline}

\begin{document}

\title{Efficient Lindblad synthesis for noise model construction}

\author{Moein Malekakhlagh}
\affiliation{IBM Thomas J. Watson Research Center, 1101 Kitchawan Rd, Yorktown Heights, NY, 10598, USA}
\author{Alireza Seif}
\affiliation{IBM Thomas J. Watson Research Center, 1101 Kitchawan Rd, Yorktown Heights, NY, 10598, USA}
\author{Daniel Puzzuoli}
\affiliation{IBM Quantum, IBM Canada, 750 West Pender St, Vancouver, BC, V6C 2T8, Canada}
\author{Luke C. G. Govia}
\affiliation{IBM Quantum, Almaden Research Center, San Jose, CA, 95120, USA}
\author{Ewout van den Berg}
\affiliation{IBM Thomas J. Watson Research Center, 1101 Kitchawan Rd, Yorktown Heights, NY, 10598, USA}

\begin{abstract}
Effective noise models are essential for analyzing and understanding the dynamics of quantum systems, particularly in applications like quantum error mitigation and correction. However, even when noise processes are well-characterized in isolation, the effective noise channels impacting target quantum operations can differ significantly, as different gates experience noise in distinct ways. Here, we present a noise model construction method that builds an effective model from a Lindbladian description of the physical noise processes acting simultaneously to the desired gate operation. It employs the Magnus expansion and Dyson series, and can be utilized for both low-order symbolic and high-order numerical approximations of the noise channel of a multi-qubit quantum gate. We envision multiple use cases of our noise construction method such as (i) computing the corresponding noise channel from a learned Lindbladian, and (ii) generating the noise channel starting with physically motivated Lindbladians for a given hardware architecture. In doing so, we close the gap between physical Lindbladians and operational level noise model parameters. We demonstrate a strong agreement between our symbolic noise construction and full numerical Lindblad simulations for various two-qubit gates, in isolation and in three- and four-qubit scenarios, for a variety of physically motivated noise sources. Our symbolic construction provides a useful breakdown of how noise model parameters depend on the underlying physical noise parameters, which gives qualitative insight into the structure of errors. For instance, our theory provides insight into the interplay of Lindblad noise with the intended gate operations, and can predict how local Lindblad noise can effectively spread into multi-qubit error. 
\end{abstract}
\date{\today}

\maketitle
\section{Introduction}
\label{Sec:Intro}

With the advent of large-scale quantum devices of hundreds of qubits, it is crucial to have scalable and easy-to-interpret quantum noise models. For instance, several quantum error mitigation (QEM) \cite{Temme_Error_2017, Li_Efficient_2017, Endo_practical_2018, Cai_Quantum_2023} methods, which are central in advancing near-term quantum applications \cite{Kandala_Error_2019, Kim_Scalable_2023, Berg_Probabilistic_2023, Kim_Evidence_2023}, require a precise model for circuit noise. Examples of such methods include probabilistic error cancellation (PEC) \cite{Temme_Error_2017, Berg_Probabilistic_2023, Gupta_Probabilistic_2023}, zero-noise extrapolation (ZNE) \cite{Temme_Error_2017, Kandala_Error_2019} in conjunction with probabilistic error amplification (PEA) \cite{Kim_Evidence_2023}, and tensor error mitigation (TEM) \cite{Filippov_Scalable_2023, Fischer_Dynamical_2024}. Meanwhile, for fault-tolerant quantum computing, accurate noise models are expected to play a role in high-performance decoders for quantum error correction (QEC) \cite{Leung_Approximate_1997, Fletcher_Optimum_2007, Nickerson_Analysing_2019, Schwartzman_Modeling_2024}, but can also contribute to architecture-tailored code choice \cite{Tuckett_Tailoring_2019, Puri_Bias_2020, Sahay_High_2023}.

\begin{figure*}[t!]
\centering
\includegraphics[scale=0.0890]{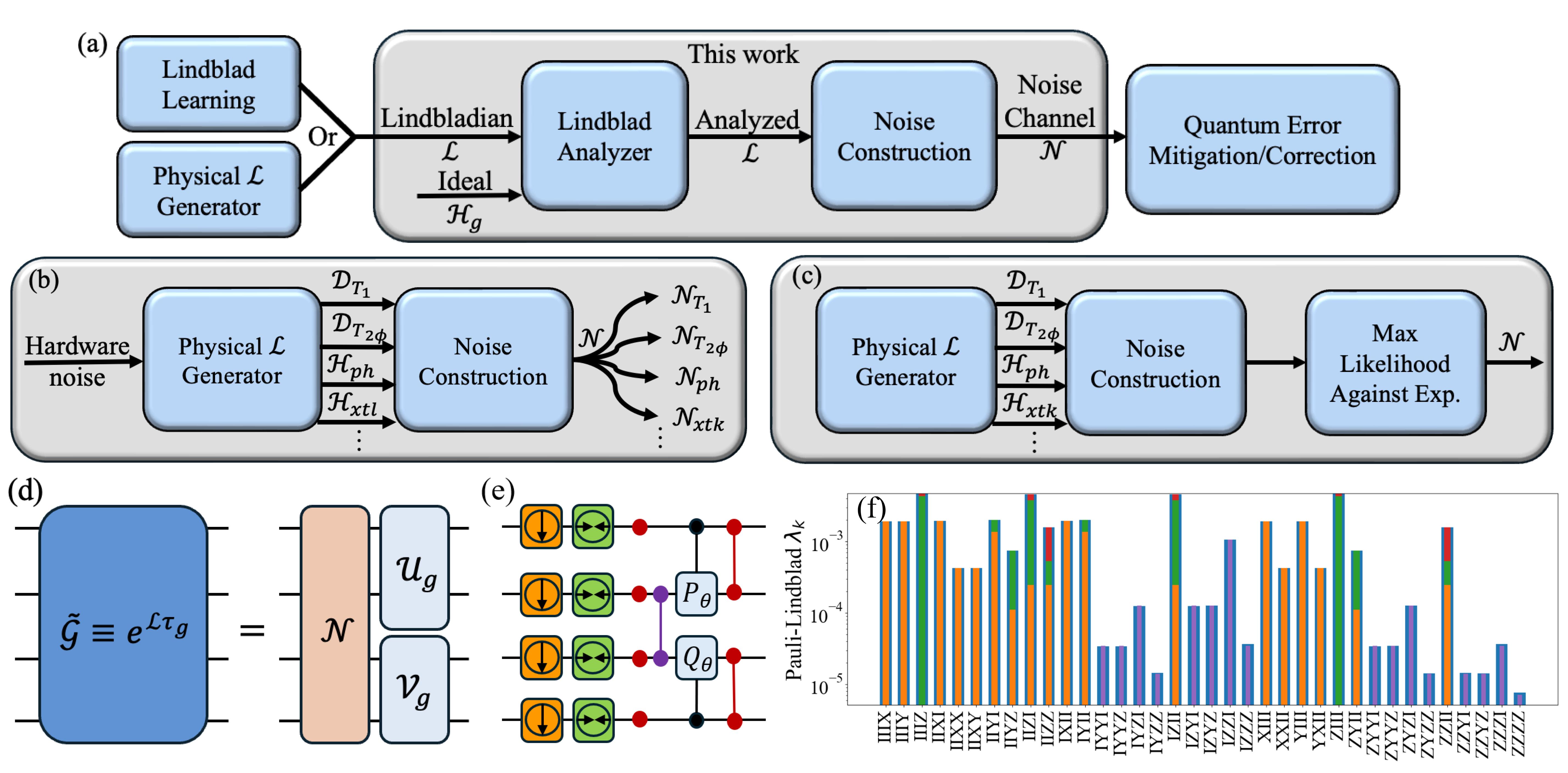}
\caption{\textbf{Schematic summary} -- (a) Proposed workflow for Lindblad noise modeling in which an input Lindbladian $\mathcal{L}$, either learned or generated with physical considerations, is fed into our proposed noise construction module in order to compute the corresponding noise channel $\mathcal{N}$. The contributions in $\mathcal{L}$ are analyzed by (i) separating the ideal part from noise via an interaction-frame representation, and (ii) categorizing coherent and incoherent terms in orders of locality (Pauli weight). We then develop a perturbative Lindblad noise construction tool, based on Magnus expansion and Dyson series, to compute the noise channel in orders of noise strength. Depending on the choice of the error mitigation/correction method, the noise construction step can compute the full channel $\mathcal{N}$ or the twirled noise (generator) as $\mathcal{N}_{\text{PL}}\equiv \exp(\mathcal{L}_{\text{PL}})$ (for PEC/PEA). We envision other applications of our noise construction method in (b) and (c). (b) Instead of learning a Lindbladian, a physics-inspired Lindblad model consisting of e.g., coherent phase and crosstalk noise and incoherent relaxation and dephasing noise, etc, can be \textit{generated} with parameters obtained from quantum hardware specifications. We find that, for weak noise, the aggregate channel (generator) is well described by the product (sum) of the individual channels (generators) for each physical noise mechanism. (c) Hardware noise parameters can potentially drift. An alternative is to use our noise construction to come up with a fit for the noise channel. The most likely parameters are then determined via comparison to experimental data. (d) A noisy layer with a noisy Lindblad channel $\tilde{\mathcal{G}}\equiv \exp(\mathcal{L}\tau_g)$ can be separated into ideal adjacent two-qubit operations and a noise channel $\mathcal{N}$, which we want to characterize. (e) A schematic 4-qubit noisy Lindblad circuit with two two-qubit gates (light blue) and added amplitude damping (orange), pure dephasing (green), coherent phase noise consisting of $Z$  and $ZZ$ terms for each gate (red), and inter-gate coherent $ZZ$ error (purple). Here, the Lindblad noise is acting \textit{continuously} and the position of the individual mechanisms is irrelevant. (f) An example of the corresponding 4-qubit Pauli-Lindblad generator parameters $\mathcal{L}_{\text{PL}}\equiv \sum_k \lambda_k (P_k \bullet P_k^{\dag} - I\bullet I)$ showing the exact value in blue, and the corresponding breakdown due to each physical mechanism (same color as panel (e)).}
\label{fig:schematic_summary}
\end{figure*}

Quantum process tomography (QPT) is a standard method of noise model characterization \cite{Poyatos_Complete_1997, Chuang_Prescription_1997, DAriano_Quantum_2001, Mohseni_Quantum_2008, Merkel_Self_2013, Greenbaum_Introduction_2015} that provides a complete description of quantum processes. However, this method is not scalable, as it requires an exponentially large number of measurement and preparation settings \cite{Mohseni_Quantum_2008}. The use of Pauli twirling~\cite{Bennett_Purification_1996,Knill_Fault_2004,Kern_Quantum_2005,Geller_Efficient_2013,Wallman_Noise_2016} shapes the noise to an effective Pauli channel and greatly reduces the number of parameters, but retains an exponential learning complexity. Scaling noise learning beyond a small number of qubits therefore requires an even more restricted structure to the ansatz for the noise model. One example of such a model is the sparse Pauli-Lindblad (PL) noise model~\cite{Berg_Probabilistic_2023}, which further restricts the Pauli channel to follow the qubit-coupling topology of the device. 

Alternatively, assuming Markovian noise, a different class of characterization methods attempts to learn the underlying Lindblad generator for the noisy gate layer \cite{Samach_Lindblad_2022, Pastori_Characterization_2022, Franca_Efficient_2024, Olsacher_Hamiltonian_2024}. One important advantage of this approach is the ability to learn the Lindbladian of any noisy gate, including non-Clifford (fractional) gates \cite{IBM_fractionalgates_2024, Layden_Theory_2024}. Moreover, Lindblad learning can potentially identify dominant coherent and incoherent noise contributions at the hardware level. Note that, experimentally, we have access only to a noisy implementation $\tilde{\mathcal{G}}$ of each ideal gate-layer $\mathcal{U}_g$, rather than an effective noise channel description (e.g.,~$\mathcal{N} = \mathcal{U}_g^\dagger\tilde{\mathcal{G}}$). 

Given a Lindbladian description of a noisy gate layer, the remaining challenge is to compute the corresponding effective noise channel needed for the QEM method or QEC application of choice. Motivated by this, we develop methods for the efficient construction of effective noise channels. Specifically, we address the following problem: How can one efficiently compute the noise channel $\mathcal{N}$ given a noisy multi-qubit quantum operation, described by a Lindblad operator $\mathcal{L}$ consisting of an ideal Hamiltonian $\mathcal{H}_g$ plus a coherent noise Hamiltonian $\mathcal{H}_{\delta}$ and an incoherent dissipator $\mathcal{D}_{\beta}$? It is worth noting that separating the noise from the ideal evolution in the Lindblad framework is \textit{not} trivial, as the standard (exact) noise construction ($\mathcal{N} = \mathcal{U}_g^\dagger\tilde{\mathcal{G}}$) requires simulating the full noisy operation, making it computationally inefficient. As an alternative, we introduce a perturbative Lindblad noise construction that allows for a controlled prediction of the noise channel, and provides important insights about the noise structure. For instance, our construction allows us to separate coherent and incoherent contributions in the effective noise channel. The former can be later suppressed using techniques such as compiler-based error compensation~\cite{Lao_Software_2022, Seif_Suppressing_2024}. A key factor in enabling our accurate noise construction is identifying a suitable frame to describe the dynamics. We find that the frame of the ideal gate \cite{Haeberlen_Coherent_1968, Mehring_Principles_2012} provides a natural separation of weak noise from strong gate energy scales required for the successful application of our perturbation theory. We develop time-dependent Lindblad perturbation theories in terms of Magnus expansion \cite{Magnus_Exponential_1954, Blanes_Magnus_2009, Blanes_Pedagogical_2010, Puzzuoli_Algorithms_2023} and Dyson series \cite{Dyson_SMatrixQED_1949, Dyson_Radiation_1949, Shillito_Fast_2021, Puzzuoli_Algorithms_2023} for both numerical and symbolic noise estimation. We demonstrate the effectiveness of our method through its strong agreement to exact numerical simulations for various noise mechanisms in multi-qubit circuits.    

We anticipate that our noise construction method will be valuable in multiple contexts starting from either (i) a learned Lindbladian, or (ii) a physically motivated Lindbladian with a small number of parameters, describing the noisy evolution of a quantum operation. Under application (i), our noise construction method serves as an essential intermediate module that takes the corresponding learned noisy Lindbladian and outputs an approximate noise channel as shown in Fig.~\ref{fig:schematic_summary}(a). Besides such a generic (physics-agnostic) application of our method, we envision other physics-inspired use cases in which the starting Lindblad model is \textit{not} directly learned, but generated based on our knowledge of the underlying hardware noise. Figure~\ref{fig:schematic_summary}(b) shows a possible implementation, where physically motivated noise is adopted from hardware specification, or measured on the fly, to construct a minimal physical noise model. Given potential drifts in hardware noise parameters \cite{Kim_Error_2024}, a more practical variation is to adopt the same physical form, but derive the most likely noise parameters through maximum likelihood estimation against experimental measurements \cite{Samach_Lindblad_2022, Tripathi_Deterministic_2024} (Fig.~\ref{fig:schematic_summary}(c)). Despite the fact that our Magnus- and Dyson-based perturbations can in principle handle pulse-level time-dependent Lindbladians, in this work we adopt a time-independent Lindblad noise model as our starting point. This choice is motivated in part by standard Lindblad learning methods \cite{Pastori_Characterization_2022, Franca_Efficient_2024, Olsacher_Hamiltonian_2024} that infer an effective (end-to-end) time-independent Markovian Lindblad generator for a given quantum operation.

We note that our proposed noise construction method can predict the full noise channel for a given Lindbladian description of a quantum operation. In various applications such as PEC/PEA, however, the corresponding Pauli-twirled noise channel is also of interest. In our symbolic analysis, we have therefore focused mainly on the Pauli-twirled noise channel $\mathcal{N}_{\text{PL}}=\exp(\mathcal{L}_{\text{PL}})$, referred to as the PL noise model, which can be expressed as the evolution of an effective diagonal dissipative Lindbladian \cite{Berg_Probabilistic_2023, Blume_Taxonomy_2022}
\begin{align}
\mathcal{L}_{\text{PL}}(\rho) = \sum_{k\in\mathcal{K}} \lambda_k (P_k \rho P_k^{\dag} - \rho) \;.
\end{align}
Here, $\mathcal{K}$ represents a selected set of Pauli operators $P_k$ with associated weights $\lambda_k$. In practice, the model ansatz is chosen such that all $P_k$ are low-weight Pauli operators with a local support that follows the qubit topology, resulting in a scalable noise model. Using our noise construction method, however, we can revisit the sparsity assumptions by deriving the noise parameters $\lambda_k$ from first principles. Our analysis provides valuable insight into the PL noise model structure and its dependence on the starting Lindbladian noise parameters.

The rest of this paper is organized as follows. In Sec.~\ref{Sec:MainResults}, we summarize our key results and observations about the origin and structure of effective PL noise model considering various coherent and incoherent Lindblad noise mechanisms. In Sec.~\ref{Sec:Model}, we review the Lindblad model as the starting point for our noise construction method motivated by Lindblad learning methods \cite{Samach_Lindblad_2022, Pastori_Characterization_2022, Franca_Efficient_2024, Olsacher_Hamiltonian_2024}. In Sec.~\ref{Sec:PertNoiseCon}, we describe our proposed Lindblad noise construction method based on the Magnus \cite{Magnus_Exponential_1954, Blanes_Magnus_2009, Blanes_Pedagogical_2010, Puzzuoli_Algorithms_2023} and Dyson \cite{Dyson_SMatrixQED_1949, Dyson_Radiation_1949, Shillito_Fast_2021, Puzzuoli_Algorithms_2023} expansions and show its utility in both numerical and analytical estimation of gate noise. Lastly, in Sec.~\ref{Sec:PhysNoise}, we close the gap between physically motivated noise mechanisms and operational noise models, and show that, to leading-order, noise models for gate layers can be constructed by combining noise models due to individual Lindblad noise mechanisms. 

\section{Summary of key observations}
\label{Sec:MainResults}

We apply our noise construction method to physically motivated \textit{continuous-time} incoherent and coherent noise mechanisms such as amplitude damping, pure dephasing, phase noise, and crosstalk in various two-, three- and four-qubit circuits and provide leading-order perturbative expressions for the effective PL noise model parameters (Tables~\ref{Tab:MagnusPert-T1Decay}--\ref{Tab:MagnusPert-4QZZCrosstalk}). From our perturbative analysis, we arrive at some useful observations on the nature of the PL noise and its transmutation:
\begin{enumerate}
\item[(i)] \textit{Origin of the PL noise model} -- This model was originally introduced \cite{Berg_Probabilistic_2023} as a fit to the generator of measured twirled noise. Here, starting from a given Lindbladian, we arrive at an effective PL noise model perturbatively from first principles. We therefore close the gap between operational noise models and the physical origin of the noise.    

\item[(ii)] \textit{Physical noise breakdown for the PL generator} -- Under a weak noise assumption, where a leading-order perturbation is sufficiently precise, we find that the PL generator, parameters $\lambda_k$, can be well described as the sum of generators due to each individual noise source, enabling a precise breakdown in terms of the underlying physical noise mechanisms. We attribute the effectiveness of the leading-order perturbation, and the corresponding noise breakdown, in part to our noise construction in the interaction frame in which the physical noise is exactly transformed by the ideal gate and the perturbation is automatically expressed in powers of weak noise-to-gate interaction ratios. Examples of noise mechanisms we consider are $T_1$ decay, pure dephasing $T_{2\phi}$, and intra- and inter-gate coherent crosstalk. Each physical noise mechanism can be analyzed once and stitched together to construct more involved noise scenarios. We show an excellent agreement between aggregate models from our lowest-order analytical results and full numerical simulation of the noise (Fig.~\ref{fig:schematic_summary}(e)--(f) and Figs.~\ref{fig:PhysNoise-2QCompToNum}--\ref{fig:PhysNoise-XtalkCompToNum}).	

\item[(iii)] \textit{Interplay between noise locality and ideal gates} -- The noise is best described in the frame of the ideal (gate) operation and, depending on how a given noise source and the ideal Hamiltonian commute, the locality of the noise generator changes. For instance, local (weight-1) physical $T_1$ and $T_{2\phi}$ noise can lead to effective weight-2 PL parameters (Tables~\ref{Tab:MagnusPert-T1Decay}--\ref{Tab:MagnusPert-T2Decay} and Figs.~\ref{fig:PhysNoise-2QCompToNum}). Furthermore, $ZZ$ crosstalk between two adjacent gates can lead to weight-3 and weight-4 PL parameters depending on the nature (role) of the gate (qubits) (Tables~\ref{Tab:MagnusPert-3QZZCrosstalk}--\ref{Tab:MagnusPert-4QZZCrosstalk} and Fig.~\ref{fig:PhysNoise-ZZCrosstalkCases}--\ref{fig:PhysNoise-XtalkCompToNum}). 

\item[(iv)] \textit{Balanced spreading of incoherent noise} -- One important property of the effective PL noise model due to incoherent noise is that the spreading of noise between weight-1 and weight-2 PL generator parameters occurs in a balanced way such that $\sum_k\lambda_k$ remains invariant (Tables~\ref{Tab:MagnusPert-T1Decay}--\ref{Tab:MagnusPert-T2Decay}). In the case of amplitude damping, regardless of the nature and the rotation axis of the operation, we find $\sum_k\lambda_k = \sum_j \beta_{\downarrow j}\tau_g/2$. Similarly, for pure dephasing noise the sum simplifies to $\sum_k\lambda_k= \sum_j \beta_{\phi j}\tau_g/2$. Here, $\beta_{\downarrow j}$ and  $\beta_{\phi j}$ are the relaxation and pure dephasing rates for qubit $j$, respectively. This means that although the fidelity of individual Pauli operators is gate-dependent, the average incoherent gate fidelity is invariant and independent of the nature of the gate. More explicitly, the resulting PL channel $\mathcal{N}_{\text{PL}}=\exp[\sum_k \lambda_k (P_k \bullet P_k^{\dag} - I \bullet I)]$ can be expressed at the leading order as $(1-\sum_k\lambda_k)I\bullet I + \sum_k \lambda_k P_k \bullet P_k^{\dag} + O(\lambda_k^2)$. The average gate fidelity \cite{Pedersen_Fidelity_2007} for such a Pauli Channel then depends only on $\sum_k \lambda_k$, resulting in the universal expression $\bar{F} = 1-[d/(d+1)]\sum_j (\beta_{\downarrow j}\tau_g+\beta_{\phi j}\tau_g)/2$ \cite{Abad_Universal_2022} with $d=4$ for two-qubit gates.

\item[(v)] \textit{Signatures of the underlying noise based on the behavior of the effective PL terms} -- Having a precise mapping from Lindbladian to PL noise parameters, we can identify certain physical noise mechanisms based on our observed PL generator parameters. For instance, the effective PL parameter due to each noise mechanism has a distinct dependence on the gate angle (time), which could be useful as an identifier. Generally, leading-order incoherent (coherent) noise generators exhibit a linear (quadratic) angle dependence. A sinusoidal behavior, on top of a linear/quadratic dependence, is a signature of non-commutativity of the noise with the ideal gate and its mixing among various Pauli terms (Tables~\ref{Tab:MagnusPert-T1Decay}--\ref{Tab:MagnusPert-4QZZCrosstalk} and Appendix~\ref{App:AngleDepofPLGen}). Moreover, assuming sufficiently weak $T_1$, $T_{2\phi}$, phase ($IZ$, $ZI$, $ZZ$), and crosstalk errors, certain PL parameters should be zero up to the lowest order. Therefore, observing non-zero values for these Pauli indices is a signature of more involved noise.

\item[(vi)] \textit{Crosstalk errors appear in unique PL indices} -- As a sub-category of item (iv), we note that having crosstalk between adjacent two-qubit gates leads to unique PL indices that do not mix with contributions from other physically motivated noise sources (Fig.~\ref{fig:PhysNoise-XtalkCompToNum}). This can serve as a unique identifier of inter-gate crosstalk. Moreover, our analysis of inter-gate crosstalk reveals non-zero weight-3 and weight-4 PL parameters not considered in former sparse (weight-2) PL models. By pinning down the non-zero higher-weight PL terms, our work generalizes the sparse PL model in an efficient manner.   
\end{enumerate}

In summary, our noise construction method lays the foundation for developing scalable noise-stitching strategies assuming local Lindblad noise and structured circuits. In particular, our leading-order generation of the PL noise model for various local Lindblad noise mechanisms in few-qubit circuits turns out to be a very effective tool for predicting the PL noise model for more complex circuits, by leveraging the simpler cases as the primary building blocks. We expect that such noise-stitching algorithms will significantly enhance noise modeling for structured quantum circuits, particularly those composed of layers with inherent symmetries, such as a tiled arrangement of neighboring two-qubit gates \cite{Berg_Probabilistic_2023}.

\section{Model}
\label{Sec:Model}

We model a noisy multi-qubit gate operation in terms of a Lindblad master equation \cite{Gorini_completely_1976, Lindblad_Generators_1976} as
\begin{align}
\begin{split}
 \dot{\rho}(t) = \mathcal{L} \rho(t)  \equiv - i [H_g  + H_{\delta},\rho(t)] + \mathcal{D}_{\beta}\rho(t)  \;,
\end{split}
\label{eq:Model-lindblad 1} 
\end{align}
where $H_g$ is the Hamiltonian for the intended operation, $H_{\delta}$ is the coherent noise Hamiltonian, and $\mathcal{D}_{\beta}$ denotes the dissipator superoperator (incoherent noise). Alternatively, by expanding over an orthonormal multi-qubit Pauli basis, we can re-express Eq.~(\ref{eq:Model-lindblad 1}) as  
\begin{align}
\begin{split}
 \dot{\rho}(t) = - i [\sum\limits_j (\omega_j+\delta_j) P_j,\rho(t)] + \sum\limits_{jk} \beta_{jk} \mathcal{D}[P_j,P_k]\rho(t)  \;,
\end{split}
\label{eq:Model-lindblad 2} 
\end{align}
with real-valued $\omega_j$ and $\delta_j$ as the ideal and noise Hamiltonian coefficients, denoted overall as $\alpha_j \equiv \omega_j+\delta_j$, and $\beta_{jk}$ as the dissipative coefficients corresponding to a generalized (non-diagonal) dissipator superoperator (see also Appendix~\ref{App:LindModel})  
\begin{align}
& \mathcal{D}[P_j,P_k]\rho(t) \equiv P_j \rho(t) P_k^{\dag} - \tfrac{1}{2} \{P_k^{\dag} P_j,\rho(t) \}\;.
\label{eq:Model-Def of D[Pi,Pj]} 
\end{align}
To describe a physical Lindblad master equation, the dissipator matrix $\bm{\beta}$ must be positive semi-definite \cite{Gorini_completely_1976, Lindblad_Generators_1976, Breuer_Theory_2002}. 

One motivation for our adoption of a time-independent Lindblad model comes from Lindbladian learning methods \cite{Pastori_Characterization_2022, Franca_Efficient_2024, Olsacher_Hamiltonian_2024} that estimate the parameters of a Lindblad noise generator in the form of Eqs.~(\ref{eq:Model-lindblad 2})--(\ref{eq:Model-Def of D[Pi,Pj]}). Through repeated application of the given noisy gate, these methods construct a linear system of equations between a set of observables and the Lindbladian coefficients. In particular, the evolution of observable $O$ under Eqs.~(\ref{eq:Model-lindblad 2})--(\ref{eq:Model-Def of D[Pi,Pj]}) can be expressed as 
\begin{align}
\begin{split}
\dot{\braket{O}} & = \sum\limits_{j} i\alpha_j \braket{[P_j,O]} \\
& + \sum\limits_{jk}  \frac{\beta_{jk}}{2} ( \braket{ P_k^{\dag} [O,P_j]} + \braket{[P_k^{\dag},O]P_j}) \;.
\end{split}
\label{eq:Model-<O> evolve}
\end{align}
Initialization and measurement in sufficient number of Pauli bases provides a full rank system of equations of the form $\dot{\braket{\bm{O}}}=\bm{M} \bm{c}$ with matrix $\bm{M}$ encoding the expectation values on the right-hand side, and the vector of Lindblad coefficients $\bm{c}\equiv [\bm{\alpha}|\bm{\beta}]^{T}$. A recent learning technique \cite{Franca_Efficient_2024} proposes to fit the evolution of low-weight observables to low-degree polynomials and estimate the derivative $\dot{\braket{O}}$ at time (depth) zero. One advantage of such depth-zero learning method is that the matrix $\bm{M}$ can be determined classically. A potential drawback of using the derivative only at depth zero is the lack of robustness to state preparation (SP) error. 

\section{Perturbative noise construction}
\label{Sec:PertNoiseCon}

Given a Lindbladian description of the dynamics of a multi-qubit operation, we develop a scheme to construct the corresponding noise channel. Importantly, the learning protocol produces the \textit{full} Lindbladian consisting of the ideal Hamiltonian in addition to coherent and incoherent noise, as shown in Eq.~(\ref{eq:Model-lindblad 2}). We emphasize that the natural frame to extract the noise is the interaction frame with respect to the ideal Hamiltonian. In what follows, we first discuss the role of interaction-frame representation in noise extraction. Next, we employ Magnus and Dyson perturbation theories to construct the noise channel in a controlled manner.  

\subsection{Noise extraction}
\label{SubSec:NoiseExtr}

Interaction-frame representations provide a natural separation of energy scales into strong gate interaction and weak noise contributions. Depending on the choice of the unitary frame transformation, various decompositions are possible in which the noise could be decomposed on the left, the right, or the middle of ideal operations (see Appendix~\ref{App:IntFrameRep}). We follow the standard noise decomposition as $\tilde{\mathcal{G}} \equiv \mathcal{U}_g \mathcal{N}$ where $\tilde{\mathcal{G}}$, $\mathcal{N}$ and $\mathcal{U}_g$ are the noisy operation, the noise, and the ideal unitary operation, respectively.

Separating the Hamiltonian as $H = H_g + H_{\delta}$ into the ideal $H_g$ and the noise part $H_{\delta}$, with Pauli decomposition $H_{\delta} \equiv \sum_j \delta_j P_j$, we employ the standard definition of the interaction frame representation as
\begin{align}
&P_{jI}(t) \equiv e^{+i H_g t} P_{j} e^{-i H_g t} \;,
\label{eq:PertNoiseCon-Def of tilde(P)_i} \\
&\rho_{I}(t) \equiv e^{+i H_g t} \rho(t) e^{-i H_g t} \;,
\label{eq:PertNoiseCon-Def of tilde(rho)}
\end{align}
where $\rho_{I}(t)$ and $P_{jI}(t)$ denote the transformed density matrix and the $j$th Pauli operator, respectively. The transformed density matrix therefore evolves according only to the noise 
\begin{align}
\begin{split}
& \dot{\rho}_I(t) = \mathcal{L}_I \rho_I(t) \equiv  -i\sum\limits_j \delta_j [P_{jI}(t),\rho_I(t)]  \\
&+ \sum\limits_{jk} \beta_{jk}\left(P_{jI}(t) \rho_I(t) P_{kI}^{\dag}(t) - \frac{1}{2}\{P_{kI}^{\dag}(t)P_{jI}(t),\rho_I(t)\}\right) \;.
\end{split}
\label{eq:PertNoiseCon-lindblad 3}
\end{align}

Under this definition of the interaction frame, the overall time evolution operator takes the form
\begin{align}
\underbrace{e^{\mathcal{L} \tau_g}}_{\text{Noisy}} = \underbrace{\mathcal{U}_g(\tau_g)}_{\text{Ideal}} \underbrace{\mathcal{T} e^{\int_{0}^{\tau_g} dt'\mathcal{L}_I(t')}}_{\text{Noise}} \;,
\label{eq:PertNoiseCon-int rep 1}
\end{align}
which is consistent with the standard circuit decomposition $\tilde{\mathcal{G}} \equiv \mathcal{U}_g \mathcal{N}$ mentioned above. Here, $\mathcal{U}_g(\tau_g)\equiv \exp(-i \mathcal{H}_g \tau_g)$ is the ideal operation, $\tau_g$ is the operation time, and $\mathcal{T}$ denotes the time-ordering operator. To compute the time-ordered noisy evolution, we develop a Lindblad perturbation theory discussed in the following.   

\subsection{Lindblad Perturbation}
\label{SubSec:LindPert}

The time-dependent nature of the interaction-frame Lindbladian~(\ref{eq:PertNoiseCon-lindblad 3}) necessitates a time-dependent Lindblad perturbation \cite{Malekakhlagh_Time_2022} method. To this end, we employ the Magnus expansion \cite{Magnus_Exponential_1954, Blanes_Magnus_2009, Blanes_Pedagogical_2010, Puzzuoli_Algorithms_2023}, which computes an effective generator for the time evolution, and the corresponding Dyson series \cite{Dyson_SMatrixQED_1949, Dyson_Radiation_1949, Shillito_Fast_2021, Puzzuoli_Algorithms_2023}. Although Magnus expansion is well studied for Hamiltonian dynamics, its application for Lindbladian evolution is less common \cite{Dai_Floquet_2016, Schnell_High_2021, Mizuta_Breakdown_2021}. However, we find its usage in the Lindbladian context very insightful. First, by computing an effective generator, Magnus is in principle consistent with error mitigation protocols that are based on quasi-probabilistic implementation of the noise generator \cite{Temme_Error_2017, Berg_Probabilistic_2023}. Second, the method is to some extent structure preserving as it preserves the trace and the Hermiticity, but not necessarily the positivity of the density matrix \cite{Haddadfarshi_Completely_2015, Schnell_High_2021}. This means that in general our method will not produce completely-positive quantum channels for the effective error. However, for the physical scenarios examined in this work, we find that, up to second order, the effective error channels remain completely-positive, and the Pauli-twirled variants of our results are completely-positive by construction.

\begin{figure}
\centering
\includegraphics[scale=0.365]{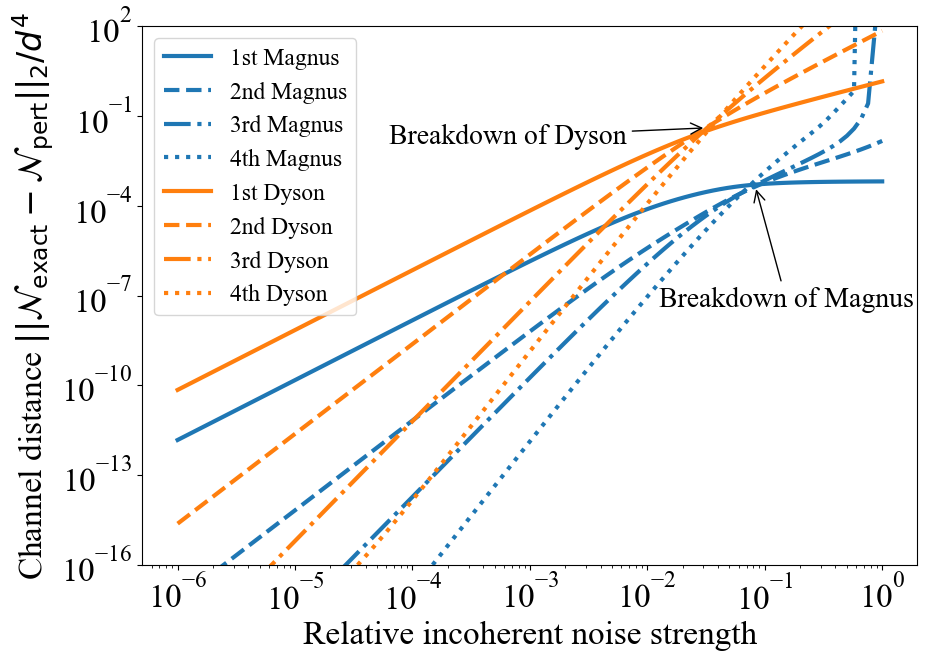}
\caption{\textbf{Numerical estimation of noise via Magnus and Dyson perturbation} -- Successive perturbative estimation (up to the fourth order) of the noise for a two-qubit $CX_{\pi/2}$ gate. We consider an ideal Hamiltonian of the form $H_g = (\omega_{cx}/2)(IX-ZX)$, such that $\omega_{cx}\tau_g = \pi/2$, and add only incoherent noise in the form of a dense random $\bm{\beta}$ matrix. The plot shows the average Frobenius distance between an exact numerical and the successive perturbative orders as $(1/d^4)||\mathcal{N}^{\text{exact}}-\mathcal{N}^{\text{pert}}||_2$, for $d=4$, as a function of the relative incoherent noise strength.}
\label{fig:NumMagnusCXpiOv4}
\end{figure}

Under the Magnus method, we compute the noise in Eq.~(\ref{eq:PertNoiseCon-int rep 1}) as
\begin{align}
\mathcal{T} e^{\int_{0}^{t} dt'\mathcal{L}_I(t')} = e^{\Omega(t,0)} \;,
\label{eq:PertNoiseCon-Def of Magnus G}
\end{align}
with $\Omega(t,0)\equiv \sum_k \Omega_k(t,0)$ as the effective noise generator, which, up to the second order, is approximated as \cite{Magnus_Exponential_1954, Blanes_Magnus_2009, Blanes_Pedagogical_2010}
\begin{align}
\Omega_1(t,0) &= \int_{0}^{t} dt'\mathcal{L}_I(t') \;,
\label{eq:PertNoiseCon-G1 Sol}\\
\Omega_2(t,0) &= \frac{1}{2} \int_{0}^{t} dt' \int_{0}^{t'} dt'' [\mathcal{L}_I(t'),\mathcal{L}_I(t'')] \;.
\label{eq:PertNoiseCon-G2 Sol} 
\end{align}
Therefore, at the leading-order, we integrate the interaction-frame Lindbladian directly. Higher-order corrections appear as multi-time integrals of nested commutators of the Lindbladian at various times. 

For symbolic calculations, taking the full matrix exponential in Eq.~(\ref{eq:PertNoiseCon-Def of Magnus G}) is only feasible for sufficiently small and sparse Lindblad noise models. Alternatively, we can further expand the time evolution operator as:
\begin{align}
\begin{split}
\mathcal{T} e^{\int_{0}^{t} dt' \mathcal{L}_I(t')} &= \mathcal{I}+\Omega_1(t,0)\\
&+\Omega_2(t,0)+\frac{1}{2}\Omega_1^2(t,0)+\mathcal{O}(\mathcal{L}_I^3)\;, 
\end{split}
\label{eq:PertNoiseCon-N_Ch Dyson Sol}
\end{align} 
which provides an unconventional, but very useful, representation of Dyson series in terms of Magnus series \cite{Burum_Magnus_1981, Salzman_Alternative_1985, Puzzuoli_Algorithms_2023}. Expressions for Magnus and Dyson perturbations up to the fourth order are provided in Appendix~\ref{App:TDLindPT}. 

We promote the use of Lindblad perturbation for noise construction in two main ways. First, it serves as a generic noise construction module that takes a learned Lindbladian, \textit{without} a consideration of its physical relevance, and outputs the resulting noise channel (Fig.~\ref{fig:schematic_summary}(a)). We have a numerical implementation of this generic use case of Magnus and Dyson up to the fourth order (see also Appendix~\ref{App:TDLindPT}). Figure~\ref{fig:NumMagnusCXpiOv4} shows the Frobenius distance of the successive orders of Magnus expansion and Dyson series from an exact computation of the noise channel for a $CX_{\pi/2}$ gate when scanning the strength of a dense random dissipator matrix. Below a relative noise strength threshold, the perturbation is convergent and higher-order corrections provide more precise estimates. In particular, the Magnus expansion demonstrates a higher threshold for convergence and also yields a lower error at a given order in comparison to the Dyson series, but this comes at a higher computational cost due to exponentiation. 

Second, starting from physically relevant coherent and incoherent noise, we employ perturbation for deriving leading-order symbolic results. Given that an exact symbolic exponentiation under the Magnus method is only possible for sufficiently simple problems, for our symbolic derivations we follow the Dyson series as in Eq.~(\ref{eq:PertNoiseCon-N_Ch Dyson Sol}). We find the commutator structure of the generators~(\ref{eq:PertNoiseCon-G1 Sol})--(\ref{eq:PertNoiseCon-G2 Sol}) to be an effective tool for an analytical description of the interplay between the underlying noise and ideal gate, and for predicting how the locality of the physical noise is transformed. This can inform and extend a given PL noise model ansatz by narrowing down what non-zero terms we expect to get for a given gate layer and a Lindblad noise mechanism, and can save resources as learning higher-weight PL fidelities becomes more expensive. In Sec.~\ref{Sec:PhysNoise}, we consider various coherent and incoherent Lindblad noise mechanisms, provide leading-order analytical expressions for the effective noise generator, and compare the result to exact numerical noise computation. 


\section{Effective Pauli-Lindblad model for physically motivated noise}
\label{Sec:PhysNoise}

We demonstrate the effectiveness of our noise construction methods from  Sec.~\ref{Sec:PertNoiseCon} by considering physically motivated coherent and incoherent noise mechanisms for two-qubit gates in multi-qubit circuits. We analyze three cases: (i) identity, (ii) controlled-X (motivated by microwave-activated cross-resonance (CR) gates \cite{Paraoanu_Microwave_2006, Rigetti_Fully_2010, Sheldon_Procedure_2016, Magesan_Effective_2020, Tripathi_Operation_2019, Malekakhlagh_First_2020, Kandala_Demonstration_2021, Malekakhlagh_Mitigating_2022, Itoko_Three_2024}), and (iii) controlled-Z (motivated by flux-activated $ZZ$ gates \cite{Yan_Tunable_2018, Foxen_Demonstrating_2020, Collodo_Implementation_2020, Sung_Realization_2021, Stehlik_Tunable_2021, Li_Realization_2024}) as our ideal two-qubit operations. In Sec.~{\ref{SubSec:IncohNoise}, we study amplitude damping ($T_1$ decay) and pure dephasing ($T_{2\phi}$ decay) for each qubit as the dominant physical incoherent noise sources. Moreover, in Sec.~\ref{SubSec:CohNoise}, we first analyze single-qubit $Z$ (Stark shift) and intra-gate $ZZ$, dubbed collectively as the phase noise, and then generalize to $ZZ$ crosstalk between a two-qubit gate and its neighboring spectator qubits as well as between two adjacent two-qubit gates.  

Despite the generality of our noise construction method, for brevity and also motivated by error mitigation models, we focus on the corresponding diagonal Pauli noise channel. Importantly, we can explicitly show that using our perturbation theory, twirling the considered physical noise mechanisms all lead to effective PL noise models of the form $\mathcal{N}_{\text{PL}} = \exp(\mathcal{L}_{\text{PL}})$ with a diagonal generator as $\mathcal{L}_{\text{PL}}\equiv \sum_k \lambda_k (P_k \bullet P_k^{\dag} - I \bullet I)$. We summarize our results in terms of leading-order effective PL $\lambda_k$ parameters for each noise source independently in Tables~\ref{Tab:MagnusPert-T1Decay}--\ref{Tab:MagnusPert-4QZZCrosstalk}. More details on the derivation of effective PL noise models are discussed in Appendix~\ref{App:WhyPauliLind} starting with more introductory examples.  

Lastly, in Sec.~\ref{SubSec:NumComp} and Figs.~\ref{fig:PhysNoise-2QCompToNum}--\ref{fig:PhysNoise-XtalkCompToNum}, we demonstrate a strong agreement between our leading-order analytical results and a numerical computation of the effective noise channel parameters. For the numerical simulations, we account for the different noise sources simultaneously. A key outcome of this comparison is the effectiveness of a physical noise breakdown for the PL noise model generator parameters $\lambda_k$, where our analytical estimates for each physical noise source (individual tables) add up very precisely to explain the numerical estimates under sufficiently weak physical noise. 

\begin{table}[t!]
\centering
\begin{tabular}{|c|c|c|c|}
\hline
gate & $I_{\tau_g}$ & $CZ_{\theta}$ & $CX_{\theta}$ \\
\hline \hline
\(\lambda_{ix}\) & $\frac{\beta_{\downarrow r}\tau_g}{4}$ & $\frac{2\theta+\sin(2\theta)}{16}\frac{\beta_{\downarrow r}}{\omega_{cz}}$ & $\frac{\theta}{4}\frac{\beta_{\downarrow r}}{\omega_{cx}}$ \\
\hline
\(\lambda_{iy}\) & $\frac{\beta_{\downarrow r}\tau_g}{4}$ & $\frac{2\theta+\sin(2\theta)}{16}\frac{\beta_{\downarrow r}}{\omega_{cz}}$ & $ \frac{12\theta + 8\sin(2\theta)+\sin(4\theta)}{128}\frac{\beta_{\downarrow r}}{\omega_{cx}}$ \\
\hline
\(\lambda_{iz}\) & 0 & 0 & $\frac{4\theta-\sin(4\theta)}{128}\frac{\beta_{\downarrow r}}{\omega_{cx}}$ \\
\hline
\(\lambda_{xi}\) & $\frac{\beta_{\downarrow l}\tau_g}{4}$ & $\frac{2\theta+\sin(2\theta)}{16}\frac{\beta_{\downarrow l}}{\omega_{cz}}$ & $\frac{2\theta+\sin(2\theta)}{16}\frac{\beta_{\downarrow l}}{\omega_{cx}}$ \\
\hline
\(\lambda_{xx}\) & 0 & 0 & $\frac{2\theta-\sin(2\theta)}{16}\frac{\beta_{\downarrow l}}{\omega_{cx}}$ \\
\hline
\(\lambda_{xz}\) & 0 & $\frac{2\theta-\sin(2\theta)}{16}\frac{\beta_{\downarrow l}}{\omega_{cz}}$ & 0 \\
\hline
\(\lambda_{yi}\) & $\frac{\beta_{\downarrow l}\tau_g}{4}$ & $\frac{2\theta+\sin(2\theta)}{16}\frac{\beta_{\downarrow l}}{\omega_{cz}}$ & $\frac{2\theta+\sin(2\theta)}{16}\frac{\beta_{\downarrow l}}{\omega_{cx}}$ \\
\hline
\(\lambda_{yx}\) & 0 & 0 & $\frac{2\theta-\sin(2\theta)}{16}\frac{\beta_{\downarrow l}}{\omega_{cx}}$ \\
\hline
\(\lambda_{yz}\) & 0 & $\frac{2\theta-\sin(2\theta)}{16}\frac{\beta_{\downarrow l}}{\omega_{cz}}$ & 0 \\
\hline
\(\lambda_{zx}\) & 0 & $\frac{2\theta-\sin(2\theta)}{16}\frac{\beta_{\downarrow r}}{\omega_{cz}}$ & 0 \\
\hline
\(\lambda_{zy}\) & 0 & $\frac{2\theta-\sin(2\theta)}{16}\frac{\beta_{\downarrow r}}{\omega_{cz}}$ & $\frac{12\theta - 8\sin(2\theta)+\sin(4\theta)}{128}\frac{\beta_{\downarrow r}}{\omega_{cx}}$ \\
\hline
\(\lambda_{zz}\) & 0 & 0 & $\frac{4\theta-\sin(4\theta)}{128}\frac{\beta_{\downarrow r}}{\omega_{cx}}$ \\
\hline
\end{tabular}
\caption{\textbf{Amplitude damping noise}--Lowest-order dependence of the PL model parameters on qubit relaxation error $\sum_{j=l,r} \beta_{\downarrow j} \mathcal{D}[S_j^{-}]$. We have considered three cases: (i) identity operation of duration $\tau_g$ ($I_{\tau_g}$), (ii) arbitrary angle controlled-X ($CX_{\theta}$) and (iii) arbitrary angle controlled-Z gates ($CZ_{\theta}$). For brevity, we only show rows with non-zero $\lambda_k$. The identity operation $I_{\tau_g}$ is parameterized in terms of operation time $\tau_g$, while the controlled operation $CZ_{\theta}$ is parameterized in terms of the rotation angle $\theta \equiv \omega_{cz}\tau_g$ (similar for $CX_{\theta}$).}
\label{Tab:MagnusPert-T1Decay}
\end{table}

\begin{table}[t!]
\centering
\begin{tabular}{|c|c|c|c|}
\hline
gate & $I_{\tau_g}$ & $CZ_{\theta}$ & $CX_{\theta}$ \\
\hline \hline
\(\lambda_{iy}\) & 0 & 0 & $\frac{4\theta - \sin(4\theta)}{64}\frac{\beta_{\phi r}}{\omega_{cx}}$ \\
\hline
\(\lambda_{iz}\) & $\frac{\beta_{\phi r}\tau_g}{2}$ & $\frac{\theta}{2} \frac{\beta_{\phi r}}{\omega_{cz}}$ & $\frac{12\theta + 8\sin(2\theta)+\sin(4\theta)}{64}\frac{\beta_{\phi r}}{\omega_{cx}}$ \\
\hline
\(\lambda_{zi}\) & $\frac{\beta_{\phi l}\tau_g}{2}$ & $\frac{\theta}{2} \frac{\beta_{\phi l}}{\omega_{cz}}$ & $\frac{\theta}{2} \frac{\beta_{\phi l}}{\omega_{cx}}$ \\
\hline
\(\lambda_{zy}\) & 0 & 0 & $\frac{4\theta - \sin(4\theta)}{64}\frac{\beta_{\phi r}}{\omega_{cx}}$ \\
\hline
\(\lambda_{zz}\) & 0 & 0 & $\frac{12\theta - 8\sin(2\theta)+\sin(4\theta)}{64}\frac{\beta_{\phi r}}{\omega_{cx}}$ \\
\hline
\end{tabular}
\caption{\textbf{Pure dephasing noise}--Lowest-order dependence of the PL model parameters on qubit dephasing error $\sum_{j=l,r} (\beta_{\phi j}/2) \mathcal{D}[Z_j]$. The considered ideal gates are the same as Table~\ref{Tab:MagnusPert-T1Decay}. For brevity, we only show rows with non-zero $\lambda_k$.}
\label{Tab:MagnusPert-T2Decay}
\end{table}

\subsection{Incoherent noise}
\label{SubSec:IncohNoise}
We first analyze incoherent error in the form of amplitude damping and pure dephasing on each qubit, in Secs.~\ref{SubSubSec:AmpDamp} and~\ref{SubSubSec:PurDeph}, respectively. We find the leading-order contribution to the PL $\lambda_k$ noise parameters due to incoherent noise  to be linear in $\beta_{\downarrow}/\omega_g$ and $\beta_{\phi}/\omega_g$, where $\beta_{\downarrow}$ and $\beta_{\phi}$ denote the relaxation and pure dephasing rates, and $\omega_{g}$ is the dominant gate interaction rate.     

\subsubsection{Amplitude damping}
\label{SubSubSec:AmpDamp}
We model amplitude damping as a continuous process via the Lindblad equation
\begin{align}
\dot{\rho}(t) = -i[H_g, \rho(t)] + \sum_{j=l,r} \beta_{\downarrow j} \mathcal{D}[S_j^{-}]\rho(t) \;,
\label{eq:PhysNoise-AmpDamp Lind}
\end{align}
where the ideal Hamiltonian $H_g$ corresponds to one of the following: (i) identity operation with $H_g = II$,  (ii) arbitrary angle $CX_{\theta}$ with $H_g=(\omega_{cx}/2)(IX-ZX)$, or (iii) arbitrary angle $CZ_{\theta}$ with $H_g=(\omega_{cz}/2)(II-IZ-ZI+ZZ)$. The gate angle $\theta$ is the effective two-qubit rotation at gate time $t=\tau_g$ defined as $\theta \equiv \omega_g \tau_g$ for $\omega_g = \omega_{cx}, \ \omega_{cz}$. $\beta_{\downarrow j}$ is the relaxation rate for qubit $j=l,r$, $S_l^{-}\equiv S^- \otimes I$ and $S_r^{-}\equiv  I\otimes S^-$ are the corresponding lowering spin operators. Under this convention, the left (right) qubit is the control (target) for the $CX_{\theta}$ operation. Moreover, the dissipator is defined as $\mathcal{D}[C](\bullet)\equiv C\bullet C^{\dag} - (1/2)(C^{\dag}C\bullet - \bullet C^{\dag}C)$.     

Following the noise construction method of Sec.~\ref{Sec:PertNoiseCon}, moving to the interaction frame with respect to $H_g$, computing the effective Magnus generator based on Eqs.~(\ref{eq:PertNoiseCon-Def of Magnus G})--(\ref{eq:PertNoiseCon-G1 Sol}), twirling the noise channel, and taking the matrix logarithm of the twirled channel, we arrive at an effective PL generator for amplitude damping noise with parameters that are summarized in Table~\ref{Tab:MagnusPert-T1Decay}, where the columns show the results for the $I_{\tau_g}$, $CZ_{\theta}$, and $CX_{\theta}$ gate operations (see Appendix~\ref{App:WhyPauliLind}).  

Under the identity operation, the noise does not transform and the non-zero PL generator parameters are the original weight-1 terms $\lambda_{ix}$, $\lambda_{iy}$, $\lambda_{xi}$ and $\lambda_{yi}$ values (left column). This is because the amplitude damping Lindbladian contains transverse projections onto the $X$ and $Y$ axes (see also Appendices~\ref{App:LindModel} and~\ref{SubApp:Id+AD+PD}). Under a non-trivial gate operation, however, different Pauli components of the noise generator can transform continuously into one another. In particular, under the $CZ_{\theta}$ gate, the non-zero PL parameters come in terms of the Pauli pairs $\lambda_{ix}\leftrightarrow\lambda_{zx}$, $\lambda_{iy}\leftrightarrow\lambda_{zy}$, $\lambda_{xi}\leftrightarrow\lambda_{xz}$, and $\lambda_{yi}\leftrightarrow\lambda_{yz}$, in which the local $T_1$ noise spreads into weight-2 terms with relative weights determined by $[2\theta\pm \sin(2\theta)]/16$ (middle column). For Clifford angles $\theta = n \pi/2$, $n \in \mathbb{Z}$, the noise is symmetrically shared between the corresponding weight-1 and weight-2 PL parameters. The noise transformation is even more involved for the $CX_{\theta}$ operation given that, unlike $CZ_{\theta}$, it has well-defined control and target roles (rightmost column). In particular, the $IX$ component of the $T_1$ noise commutes with $CX_{\theta}$, leaving $\lambda_{ix}$ unchanged. The $\lambda_{iy}$ PL parameter, however, spreads into $\lambda_{zy}$, $\lambda_{iz}$ and $\lambda_{zz}$. Moreover, the $T_1$ noise on the left qubit is also shared between $\lambda_{xi} \leftrightarrow \lambda_{xx}$ and $\lambda_{yi} \leftrightarrow \lambda_{yx}$. At Clifford angles, one again finds pairwise symmetric values up to the leading order.

\subsubsection{Pure dephasing}
\label{SubSubSec:PurDeph}
We next analyze the interplay of pure dephasing with the three considered two-qubit operations. We describe pure dephasing noise via the Lindblad equation 
\begin{align}
\dot{\rho}(t) = -i[H_g, \rho(t)] + \sum_{j=l,r} \frac{\beta_{\phi j}}{2} \mathcal{D}[Z_j]\rho(t) \;,
\label{eq:PhysNoise-AmpDamp Lind}
\end{align}
where $\beta_{\phi j}$ is the pure dephasing rate for qubit $j=l,r$, and $Z_l \equiv Z \otimes I$ and $Z_r \equiv I\otimes Z$ are the single-qubit $Z$ operators. Following similar steps as described in Sec.~\ref{SubSubSec:AmpDamp}, we compute an effective PL generator due to pure dephasing noise which is summarized in Table~\ref{Tab:MagnusPert-T2Decay}.

Given the diagonal ($Z$) form of the pure dephasing dissipator in Eq.~(\ref{eq:PhysNoise-AmpDamp Lind}), it commutes with both $I_{\tau_g}$ and $CZ_{\theta}$ operations, leading to a trivial PL model with only the original $\lambda_{iz}$ and $\lambda_{zi}$ being non-zero without any mixing (left and middle columns of Table~\ref{Tab:MagnusPert-T2Decay}). For the $CX_{\theta}$ operation, pure dephasing on the control qubit commutes with the ideal gate, leaving $\lambda_{zi}$ unchanged. Pure dephasing on the target qubit, however, spreads from the starting $\lambda_{iz}$ onto $\lambda_{iy}$, $\lambda_{zy}$ and $\lambda_{zz}$ (right column).    
  
\begin{table}[t!]
\centering
\begin{tabular}{|c|c|c|c|}
\hline
gate & $I_{\tau_g}$ & $CZ_{\theta}$ & $CX_{\theta}$ \\
\hline\hline
\(\lambda_{iy}\) & 0 & 0 & $\frac{\sin^4(\theta_{g})}{16}\frac{(\delta_{iz}-\delta_{zz})^2}{\omega_{cx}^2}$ \\
\hline
\(\lambda_{iz}\) & $\frac{\delta_{iz}^2\tau_g^2}{4}$ & $\frac{\theta_{cz}^2}{4}\frac{\delta_{iz}^2}{\omega_{cz}^2}$ & $\frac{[2\theta (\delta_{iz}+\delta_{zz}) +  \sin(2\theta_{g})(\delta_{iz}-\delta_{zz})]^2}{64 \omega_{cx}^2}$ \\
\hline
\(\lambda_{zi}\) & $\frac{\delta_{zi}^2\tau_g^2}{4}$ & $\frac{\theta_{cz}^2}{4}\frac{\delta_{zi}^2}{\omega_{cz}^2}$ & $\frac{\theta_{cz}^2}{4}\frac{\delta_{zi}^2}{\omega_{cx}^2}$ \\
\hline
\(\lambda_{zy}\) & 0 & 0 & $\frac{\sin^4(\theta_{g})}{16}\frac{(\delta_{iz}-\delta_{zz})^2}{\omega_{cx}^2}$ \\
\hline
\(\lambda_{zz}\) & $\frac{\delta_{zz}^2\tau_g^2}{4}$ & $\frac{\theta_{cz}^2}{4}\frac{\delta_{zz}^2}{\omega_{cz}^2}$ & $\frac{[2\theta (\delta_{iz}+\delta_{zz}) - \sin(2\theta_{g}) (\delta_{iz}-\delta_{zz})]^2}{64 \omega_{cx}^2}$\\
\hline
\end{tabular}
\caption{\textbf{Two-qubit phase noise}--Lowest-order dependence of the PL model parameters on Hamiltonian phase error $H_{\delta} \equiv (\delta_{iz}/2)IZ + (\delta_{zi}/2)ZI + (\delta_{zz}/2)ZZ$. We have considered three cases: (i) identity operation of duration $\tau_g$ ($I_{\tau_g}$), (ii) arbitrary angle controlled-X ($CX_{\theta}$) and (iii) arbitrary angle controlled-Z gates ($CZ_{\theta}$). For brevity, we only show rows with non-zero $\lambda_k$.}
\label{Tab:MagnusPert-2QPhaseError}
\end{table}

\begin{figure*}[t!]
\centering
\includegraphics[scale=0.70]{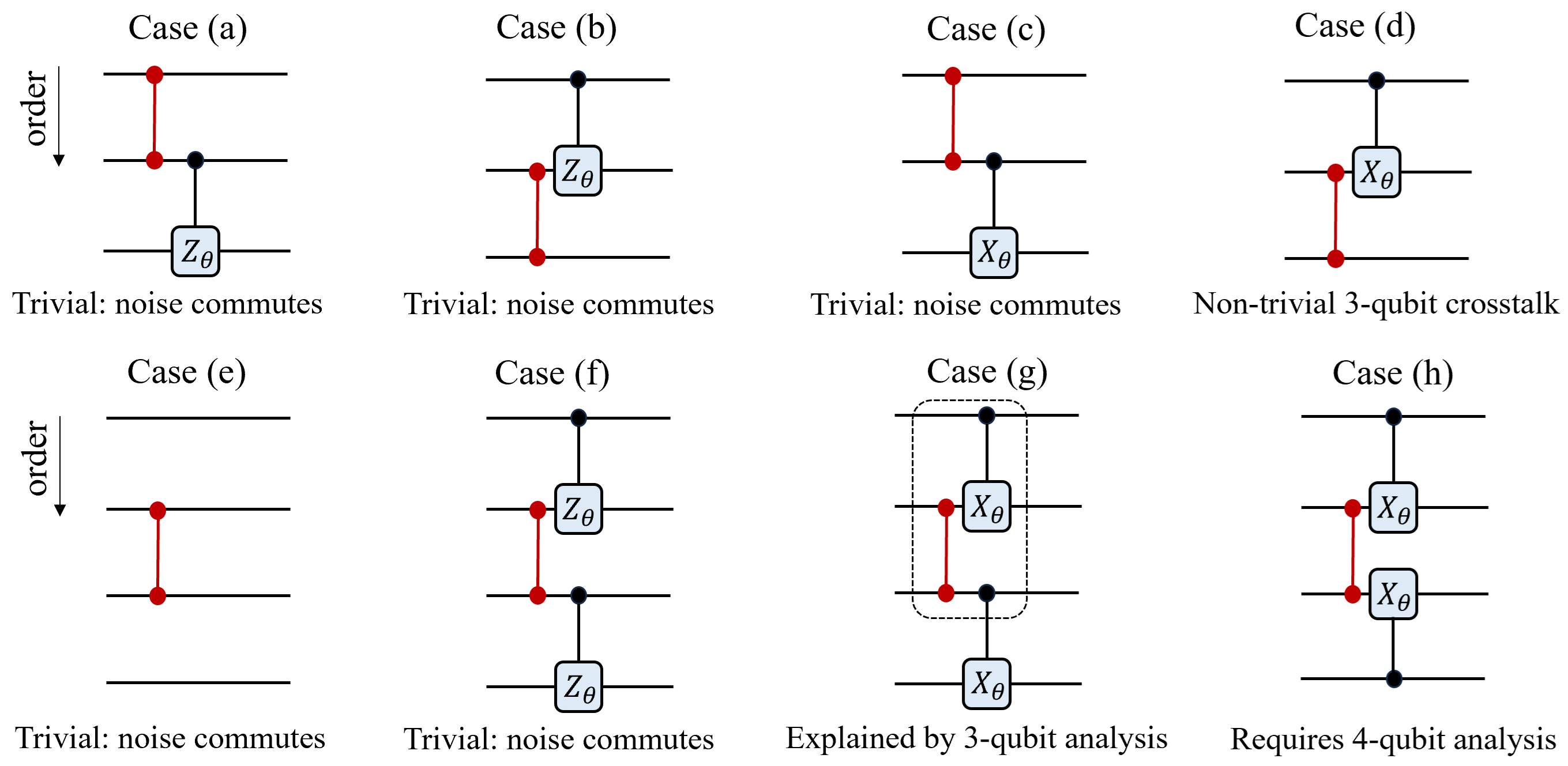}
\caption{\textbf{Various $ZZ$ crosstalk scenarios} -- Top row shows cases of $ZZ$ crosstalk between an active gate and a third spectator qubit (results summarized in Table~\ref{Tab:MagnusPert-3QZZCrosstalk}), while the bottom row summarizes $ZZ$ crosstalk cases between two adjacent two-qubit gates (results summarized in Table~\ref{Tab:MagnusPert-4QZZCrosstalk}). (a)--(b) $ZZ$ crosstalk between a $CZ_{\theta}$ gate and a control or target spectator qubit is trivial as it commutes with the ideal gate. (c)--(d) $ZZ$ between a $CZ_{\theta}$ and a control spectator is also trivial, but not with a target spectator qubit and can lead to weight-3 PL noise parameters. (e)--(f) $ZZ$ crosstalk between idle or $CZ_{\theta}$ gates are trivial. (g) $ZZ$ crosstalk between adjacent control and target qubits of two neighboring $CX_{\theta}$ gates turns out to be the same as case (d). (h) $ZZ$ crosstalk between the adjacent target qubits of neighboring two $CX_{\theta}$ gates leads to rather involved weight-3 and weight-4 PL noise parameters (right column of Table~\ref{Tab:MagnusPert-4QZZCrosstalk}). We model the crosstalk noise as a continuous process happening simultaneously with the ideal gates affected by the noise Hamiltonian $H_{\delta}= (\delta_{izzi}/2)IZZI$. Our graphical representation of the crosstalk noise as the (red) dumbbells before the ideal gates is just for better visibility.}
\label{fig:PhysNoise-ZZCrosstalkCases}
\end{figure*}

\subsection{Coherent noise}
\label{SubSec:CohNoise}

A prevalent source of coherent noise in a quantum processor is unwanted quantum crosstalk interaction, \textit{mainly} between neighboring qubits. For instance, in trapped ions, implementing a Molmer-Sorensen gate \cite{Sorensen_Quantum_1999} comes with a coherent $XX$ or $YY$ crosstalk \cite{Wu_noise_2018}. In superconducting qubits, nearest-neighbor crosstalk typically appears as a $ZZ$ interaction. This originates both through static interactions with higher qubit levels, as well as due to dynamic shift of the energy levels during gate operations. This type of noise is detrimental during two-qubit gate operation (depending on the nature of the gate), during idle times, as well as between adjacent two-qubit gates. In addition to crosstalk, there can be single-qubit $Z$ error due to gate operations (Stark shift), frame mismatch, or drift in qubit frequencies. 

Leaving the physical origin aside, motivated by superconducting architectures, in sec.~\ref{SubSubSec:2QPhaseNoise}, we study two-qubit $IZ$, $ZI$ and $ZZ$ Hamiltonian interactions, collectively referred to as the phase noise, for various two-qubit operations.  Furthermore, in Sec.~\ref{SubSubSec:QXtalk}, considering multiple crosstalk scenarios described in Fig.~\ref{fig:PhysNoise-ZZCrosstalkCases}, we study $ZZ$ quantum crosstalk between an active two-qubit gate and spectator qubits, and also between two adjacent two-qubit gates. We find that the twirled noise channel due to the coherent phase noise and inter-gate crosstalk can again be generated via an effective PL noise model. We summarize the effective PL model parameters in Tables~\ref{Tab:MagnusPert-2QPhaseError}--\ref{Tab:MagnusPert-4QZZCrosstalk}. Compared to incoherent noise, for which the leading-order contribution to the PL generator is linear in relaxation and dephasing rates, for coherent noise we find the lowest-order contributions to be quadratic in the Hamiltonian noise parameters. An important finding of our crosstalk analysis is that weight-2 coherent crosstalk can turn into effective weight-3 and weight-4 PL noise generator terms depending on the commutativity of the crosstalk and the intended gate layer.      
         
\subsubsection{Two-qubit phase noise}
\label{SubSubSec:2QPhaseNoise}
We model two-qubit coherent phase noise starting from the Lindbladian    
\begin{align}
\dot{\rho}(t) = -i[H_g + H_{\delta}, \rho(t)] \;,
\label{eq:PhysNoise-CohNoise Lind}
\end{align}
where $H_g$ is any of the three ideal gate operations as described in Sec.~\ref{SubSec:IncohNoise}, and the phase noise Hamiltonian is defined as
\begin{align}
H_{\delta} \equiv \frac{\delta_{iz}}{2} IZ + \frac{\delta_{zi}}{2}ZI + \frac{\delta_{zz}}{2}ZZ \;,
\label{eq:PhysNoise-Def of H_delta}
\end{align}
parameterized with independent single-qubit $Z$ noise rates $\delta_{iz}$ and $\delta_{zi}$, and two-qubit $ZZ$ noise rate $\delta_{zz}$. 

Following the same noise construction method, as in the case of incoherent noise, we find that the twirled noise channel is generated in terms of an effective PL noise model summarized in Table~\ref{Tab:MagnusPert-2QPhaseError}. In particular, the Hamiltonian phase noise~(\ref{eq:PhysNoise-Def of H_delta}) commutes with the $I_{\tau_g}$ and $CZ_{\theta}$ operations, leaving the individual Pauli components of the PL noise generator unchanged as $\lambda_k = (\delta_k^2\tau_g^2)/4$ for $k =iz, \ zi, \ zz$. For the $CX_{\theta}$ operation, however, only the $ZI$ noise (on the control) commutes, hence unchanged, while the $IZ$ and $ZZ$ components of the Hamiltonian spread and transform into nonzero PL generator parameters $\lambda_{iy}$, $\lambda_{zy}$ (right column of Table~\ref{Tab:MagnusPert-2QPhaseError}).    

\subsubsection{Quantum crosstalk}
\label{SubSubSec:QXtalk}

\begin{table}[t!]
\centering
\begin{tabular}{|c|c|c|c|}
\hline
gate & $I_{\tau_g}$  & $I \otimes CZ_{\theta}$ & $I \otimes CX_{\theta}$\\
\hline
$\lambda_{zzi}$ & $\frac{\delta_{zzi}^2\tau_g^2}{4}$ & $\frac{\theta_{g}^2}{4}\frac{\delta_{zzi}^2}{\omega_{cz}^2}$ & $\frac{\theta_{g}^2}{4}\frac{\delta_{zzi}^2}{\omega_{cz}^2}$ \\
\hline\hline
gate & $I_{\tau_g}$ & $CZ_{\theta}\otimes I$ & $CX_{\theta}\otimes I$\\
\hline
$\lambda_{iyz}$ & 0 & 0 & $\frac{\sin^4(\theta)}{16} \frac{\delta_{izz}^2}{\omega_{cx}^2}$ \\
\hline
$\lambda_{izz}$ & $\frac{\delta_{izz}^2\tau_g^2}{4}$ & $\frac{\theta_{g}^2}{4}\frac{\delta_{izz}^2}{\omega_{cz}^2}$ & $\frac{[2\theta+\sin(2\theta)]^2}{64} \frac{\delta_{izz}^2}{\omega_{cx}^2}$ \\
\hline
$\lambda_{zyz}$ & 0 & 0 & $\frac{\sin^4(\theta)}{16} \frac{\delta_{izz}^2}{\omega_{cx}^2}$ \\
\hline
$\lambda_{zzz}$ & 0 & 0 & $\frac{[2\theta-\sin(2\theta)]^2}{64} \frac{\delta_{izz}^2}{\omega_{cx}^2}$ \\
\hline
\end{tabular}
\caption{\textbf{Three-qubit $ZZ$ crosstalk noise}--Lowest-order dependence of the PL model parameters on spectator $ZZ$ crosstalk. Top (bottom) parts show the case of control (target) spectator. For brevity, we only show non-zero PL parameters. The only non-trivial case is having $IZZ$ crosstalk $H_{\delta} = (\delta_{izz}/2) IZZ$ for the target spectator of the $CX_{\theta}\otimes I$ operation (right column of the bottom table). For other cases, the $ZZ$ crosstalk commutes with the ideal gate, hence the noise does not mix non-trivially.}
\label{Tab:MagnusPert-3QZZCrosstalk}
\end{table}

\begin{table}[t!]
\centering
\begin{tabular}{|c|c|c|}
\hline
gate & $CX_{\theta} \otimes CX_{\theta}$ & $CX_{\theta} \otimes X_{\theta}C$ \\
\hline \hline
$\lambda_{iyyi}$ & 0 & $\frac{[4\theta - \sin(4\theta)]^2}{4096}\frac{\delta_{izzi}^2}{\omega_{cx}^2}$ \\
\hline
$\lambda_{iyyz}$ & 0 & $\frac{[4\theta - \sin(4\theta)]^2}{4096}\frac{\delta_{izzi}^2}{\omega_{cx}^2}$ \\
\hline
$\lambda_{iyzi}$ & $\frac{\sin^4(\theta)}{16} \frac{\delta_{izzi}^2}{\omega_{cx}^2}$ & $\frac{\sin^4(\theta)[3+\cos(2\theta)]^2}{256}\frac{\delta_{izzi}^2}{\omega_{cx}^2}$ \\
\hline
$\lambda_{iyzz}$ & 0 & $\frac{\sin^8(\theta)}{64}\frac{\delta_{izzi}^2}{\omega_{cx}^2}$ \\
\hline
$\lambda_{izyi}$ & 0 & $\frac{\sin^4(\theta)[3+\cos(2\theta)]^2}{256}\frac{\delta_{izzi}^2}{\omega_{cx}^2}$ \\
\hline
$\lambda_{izyz}$ & 0 & $\frac{\sin^4(\theta)[3+\cos(2\theta)]^2}{256}\frac{\delta_{izzi}^2}{\omega_{cx}^2}$ \\
\hline
$\lambda_{izzi}$  & $\frac{[2\theta+\sin(2\theta)]^2}{64} \frac{\delta_{izzi}^2}{\omega_{cx}^2}$ & $\frac{[12\theta+8\sin(2\theta)+\sin(4\theta)]^2}{4096}\frac{\delta_{izzi}^2}{\omega_{cx}^2}$ \\
\hline
$\lambda_{izzz}$ & 0 & $\frac{[4\theta - \sin(4\theta)]^2}{4096}\frac{\delta_{izzi}^2}{\omega_{cx}^2}$ \\
\hline
$\lambda_{zyyi}$ & 0 & $\frac{[4\theta - \sin(4\theta)]^2}{4096}\frac{\delta_{izzi}^2}{\omega_{cx}^2}$ \\
\hline
$\lambda_{zyyz}$ & 0 & $\frac{[4\theta - \sin(4\theta)]^2}{4096}\frac{\delta_{izzi}^2}{\omega_{cx}^2}$ \\
\hline
$\lambda_{zyzi}$ & $\frac{\sin^4(\theta)}{16} \frac{\delta_{izzi}^2}{\omega_{cx}^2}$ & $\frac{\sin^4(\theta)[3+\cos(2\theta)]^2}{256}\frac{\delta_{izzi}^2}{\omega_{cx}^2}$ \\
\hline
$\lambda_{zyzz}$ & 0 & $\frac{\sin^8(\theta)}{64}\frac{\delta_{izzi}^2}{\omega_{cx}^2}$ \\
\hline
$\lambda_{zzyi}$ & 0 & $\frac{\sin^8(\theta)}{64}\frac{\delta_{izzi}^2}{\omega_{cx}^2}$ \\
\hline
$\lambda_{zzyz}$ & 0 & $\frac{\sin^8(\theta)}{64}\frac{\delta_{izzi}^2}{\omega_{cx}^2}$ \\
\hline
$\lambda_{zzzi}$ & $\frac{[2\theta-\sin(2\theta)]^2}{64} \frac{\delta_{izzi}^2}{\omega_{cx}^2}$ & $\frac{[4\theta - \sin(4\theta)]^2}{4096}\frac{\delta_{izzi}^2}{\omega_{cx}^2}$ \\
\hline
$\lambda_{zzzz}$ & 0 & $\frac{[12\theta-8\sin(2\theta)+\sin(4\theta)]^2}{4096}\frac{\delta_{izzi}^2}{\omega_{cx}^2}$ \\
\hline
\end{tabular}
\caption{\textbf{Four-qubit ZZ crosstalk noise}--Lowest-order dependence of the PL model parameters on spectator $ZZ$ crosstalk. We consider $IZZI$ crosstalk between the middle qubits of two adjacent two-qubit gates. For brevity, we only show non-zero PL parameters. The cases of $I_{\tau_g}$ and $CZ_{\theta} \otimes CZ_{\theta}$ operations are trivial as they commute with the $IZZI$ noise, and are dropped from the table for simplicity. The case of $CX_{\theta} \otimes CX_{\theta}$ is effectively the same as the control-spectator case $CX_{\theta} \otimes I$ in Table~\ref{Tab:MagnusPert-3QZZCrosstalk}. The case of two simultaneous $CX_{\theta}$ gates with nearest-neighbor targets, i.e. $CX_{\theta}\otimes X_{\theta}C$, requires a four-qubit analysis and results in 16 non-zero PL $\lambda_k$ parameters (last column). We note that cases (e) and (f) of Fig.~\ref{fig:PhysNoise-ZZCrosstalkCases} are trivial: only the $\lambda_{izzi}$ rate is non-zero as $\lambda_{izzi} = \delta_{izzi}^2\tau_g^2/4$, and therefore omitted from this table for simplicity.}
\label{Tab:MagnusPert-4QZZCrosstalk}
\end{table}

\begin{figure*}[t!]
\centering
\includegraphics[scale=0.089]{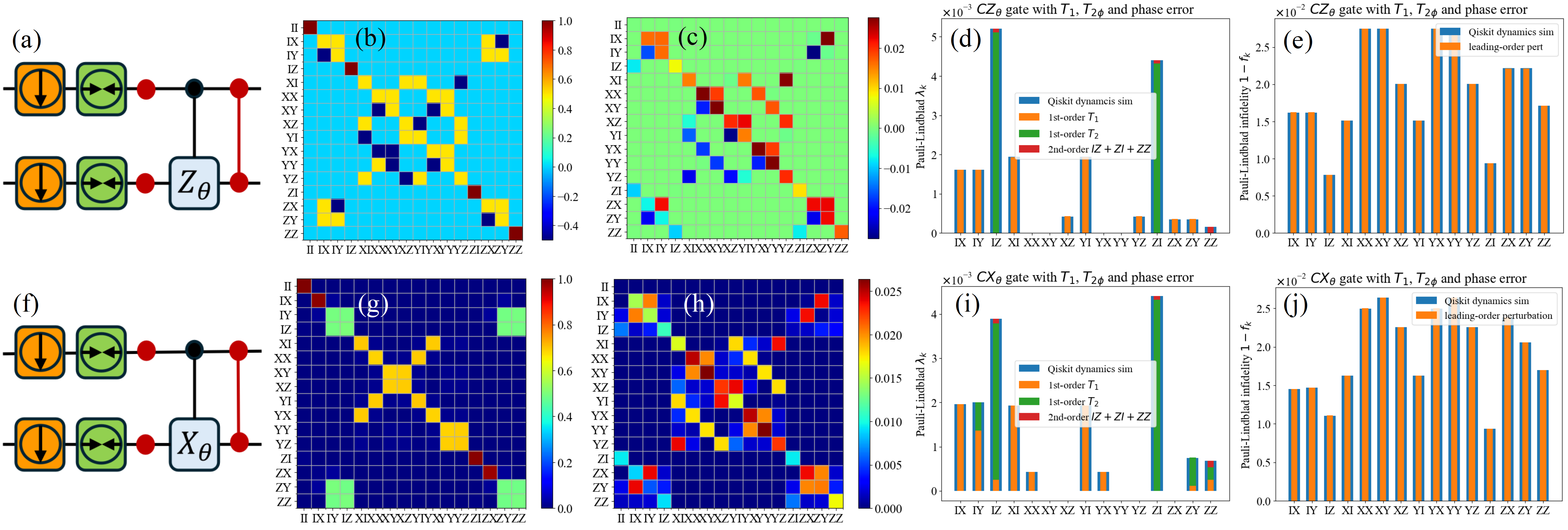}
\caption{\textbf{Comparison between leading-order perturbative PL noise model and numerical Lindblad simulation of two-qubit gate operations} -- The top and bottom rows summarize the two-qubit noise analysis for $CZ_{\pi/4}$ and $CX_{\pi/4}$ operations, respectively. Numerical results are found via simulation of Eq.~(\ref{eq:PhysNoise-FullNumLind}) using Qiskit Dynamics \cite{Puzzuoli_Qiskit_2023, Puzzuoli_Algorithms_2023} in which all noise sources are accounted for simultaneously. The perturbative noise model is constructed by adding the effective PL generator for each individual noise mechanism from Tables~\ref{Tab:MagnusPert-T1Decay}--\ref{Tab:MagnusPert-2QPhaseError}. (a) Schematic circuit for $CZ_{\theta}$ operation with \textit{continuous} $T_1$ (orange boxes), $T_{2\phi}$ (green boxes) and coherent phase noise (red dots and dumbbells) according to Eqs.~(\ref{eq:PhysNoise-FullNumLind}). (b) Numerical PTM of the Lindblad channel $\tilde{\mathcal{G}} = \exp(\mathcal{L}\tau_g)$ for $\theta=\pi/4$. (c) The noise PTM is numerically obtained as $\mathcal{N} \equiv \mathcal{U}_{{cz}_{\pi/4}}^{-1}\tilde{\mathcal{G}} \approx \mathcal{I}$ with $\mathcal{U}_{cz_{\pi/4}} \equiv \exp[-i\pi/8 (II-IZ-ZI+ZZ)]$. The plot here shows the PTM of $\mathcal{I}-\mathcal{N}$ for better visibility. (d) Parameters of PL noise generator obtained from the symplectic transformation of fidelities as $\vec{\lambda} = -(1/2)S^{-1} \log(\vec{f})$ \cite{Berg_Probabilistic_2023}, where $\vec{f}$ is the Pauli fidelities obtained from the diagonal elements of the PTM of $\mathcal{N}$ in panel (c). Perturbative estimates for each noise mechanism, color-coded according to panel (a), provide a very precise breakdown of the full numerical estimate (blue bars). (e) Comparison of perturbative and numerical Pauli fidelities $\vec{f}$ reveals a good agreement. For better visibility, here, $1-\vec{f}$ is plotted. Panels (f)--(j) show the result for the $CX_{\pi/4}$ operation in the same format and with the same noise parameters. Simulation parameters are $\omega_{g}\tau_g = \pi/4$, $\beta_{\downarrow l}/\omega_{g}=0.012$, $\beta_{\downarrow r}/\omega_{g}=0.010$, $\beta_{\phi l}/\omega_{g}=0.011$, $\beta_{\phi r}/\omega_{g}=0.013$, $\delta_{iz}/\omega_{g}=0.025$, $\delta_{zi}/\omega_{g}=0.023$, and $\delta_{zz}/\omega_{g}=0.032$ for $\omega_g = \omega_{cz}, \ \omega_{cx}$.}
\label{fig:PhysNoise-2QCompToNum}
\end{figure*}

We extend the two-qubit phase noise model of Sec.~\ref{SubSubSec:2QPhaseNoise} to account for coherent $ZZ$ crosstalk. We analyze a few physically motivated $ZZ$ crosstalk scenarios: (i) between a two-qubit gate and a third spectator qubit, and (ii) between two adjacent two-qubit gates. These two main scenarios are summarized as the top and bottom rows of Fig.~\ref{fig:PhysNoise-ZZCrosstalkCases}, respectively. Given the (approximate) additivity of noise at the level of PL noise generator, such three- and four-qubit circuits can serve as useful building blocks for the noise construction of more complex layers, which arise in quantum error mitigation or other applications. The leading-order perturbative PL model for the three-qubit and four-qubit crosstalk scenarios are summarized in Tables~\ref{Tab:MagnusPert-3QZZCrosstalk}--\ref{Tab:MagnusPert-4QZZCrosstalk}.  

Three-qubit $ZZ$ crosstalk falls into the cases of control spectator (Fig.~\ref{fig:PhysNoise-ZZCrosstalkCases}(a),(c)), which we model via a Lindblad equation of the form~(\ref{eq:PhysNoise-CohNoise Lind}) with the ideal and crosstalk Hamiltonians as
\begin{align}
& H_{g} = 
\begin{cases}
\frac{\omega_{cz}}{2}(III-IIZ-IZI+IZZ),  & \text{for} \ I \otimes CZ_{\theta},\\
 \frac{\omega_{cx}}{2}(IIX-IZX), & \text{for} \ I \otimes CX_{\theta},
\end{cases} 
\label{eq:PhysNoise-ContSpecXtalk Hg}\\
& H_{\delta} = \frac{\delta_{zzi}}{2}ZZI \;,
\label{eq:PhysNoise-ContSpecXtalk Hdelta}
\end{align}
and the case of target spectator (Fig.~\ref{fig:PhysNoise-ZZCrosstalkCases}(b),(d)), which is modeled with the Hamiltonian
\begin{align}
& H_{g} = 
\begin{cases}
\frac{\omega_{cz}}{2}(III-IZI-ZII+ZZI),  & \text{for} \  CZ_{\theta}\otimes I,\\
 \frac{\omega_{cx}}{2}(IXI-ZXI), & \text{for} \ CX_{\theta}\otimes I,
\end{cases} 
\label{eq:PhysNoise-TarSpecXtalk Hg} \\
& H_{\delta} = \frac{\delta_{izz}}{2}IZZ \;.
\label{eq:PhysNoise-TarSpecXtalk Hdelta}
\end{align}
For each case, we analyze the interplay of $CZ_{\theta}$ and $CX_{\theta}$ operations with such a crosstalk.  

\begin{figure*}[t!]
\centering
\includegraphics[scale=0.0945]{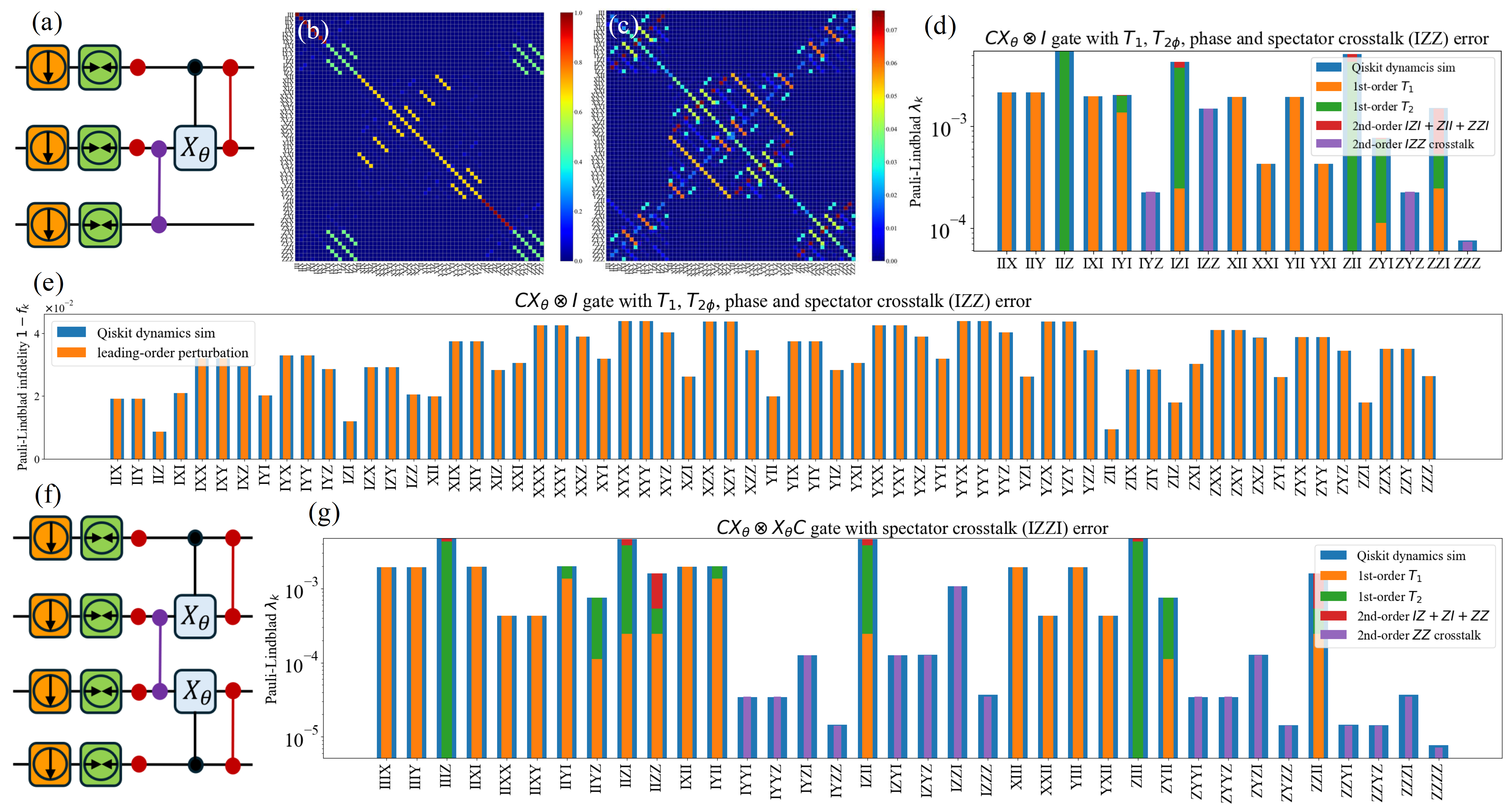}
\caption{\textbf{Comparison between leading-order perturbative PL noise model and numerical Lindblad simulation of inter-gate crosstalk}-- We consider here the non-trivial crosstalk scenarios explored in Fig.~\ref{fig:PhysNoise-ZZCrosstalkCases}(d) and  Fig.~\ref{fig:PhysNoise-ZZCrosstalkCases}(h), with a $ZZ$ crosstalk interaction acting on target qubits of $CX_{\pi/4}\otimes I$ [panels (a)--(e)] and $CX_{\pi/4}\otimes X_{\pi/4}C$ [panels (f)--(g)] operations, respectively. Numerical results are found via simulation of Eqs.~(\ref{eq:PhysNoise-TarSpecXtalk Hg})--(\ref{eq:PhysNoise-TarSpecXtalk Hdelta}) for the $CX_{\pi/4}\otimes I$ case, and Eqs.~(\ref{eq:PhysNoise-4QXtalk Hdelta}) and~(\ref{eq:PhysNoise-4QXtalk H_CXxXC}) for the $CX_{\pi/4}\otimes X_{\pi/4}C$ case, using Qiskit Dynamics \cite{Puzzuoli_Qiskit_2023, Puzzuoli_Algorithms_2023}, in which all noise sources are accounted for simultaneously. The perturbative noise model is constructed by adding the effective PL generator for each individual noise mechanism from Tables~\ref{Tab:MagnusPert-T1Decay}--\ref{Tab:MagnusPert-4QZZCrosstalk}. (a) Schematic circuit for the $CX_{\pi/4}\otimes I$ operation with \textit{continuous} $T_1$, $T_{2\phi}$, coherent phase and crosstalk noise according to Eq.~(\ref{eq:PhysNoise-FullNumLind}). (b) Numerical PTM of the Lindblad channel $\tilde{\mathcal{G}} = \exp(\mathcal{L}\tau_g)$. (c) Numerical noise PTM $\mathcal{N} \equiv \mathcal{U}_{{cx}_{\pi/4}\otimes I}^{-1}\tilde{\mathcal{G}} \approx \mathcal{I}$ with $\mathcal{U}_{cx_{\pi/4}\otimes I} \equiv \exp[-i\pi/8 (IXI-ZXI)]$, where we plot the PTM of $\mathcal{I}-\mathcal{N}$ for better visibility. (d) Parameters of the corresponding PL noise generator obtained from $\vec{\lambda} = -(1/2)S^{-1} \log(\vec{f})$ \cite{Berg_Probabilistic_2023} with Pauli fidelity $\vec{f}$ obtained from the diagonal elements of panel (c). Perturbative estimates for each noise mechanism, including $ZZ$ crosstalk (purple), color-coded according panel (a), provide a very precise breakdown of the full numerical estimate (blue bars). Here, for improved visibility of the breakdown, we use log scale and drop elements smaller than $10^{-5}$. (e) Comparison of perturbative and numerical Pauli fidelities error $1-\vec{f}$ shows very good agreement. (f) Schematic circuit for the $CX_{\pi/4}\otimes X_{\pi/4}C$ operation with similar noise process and inter-gate $ZZ$ crosstalk. (g) Due to the large number of terms for such four-qubit systems, we omitted the PTM and fidelity plots, and only show the agreement between dominant perturbative (first order) and numerical (terms approximately larger than $10^{-5}$) PL noise parameters. Simulation parameters for case (a) are $\omega_{cx}\tau_g = \pi/4$, $\beta_{\downarrow c}/\omega_{g}=0.012$, $\beta_{\downarrow t}/\omega_{g}=0.010$, $\beta_{\downarrow s}/\omega_{g}=0.011$, $\beta_{\phi c}/\omega_{g}=0.011$, $\beta_{\phi t}/\omega_{g}=0.013$, $\beta_{\phi s}/\omega_{g}=0.014	$, $\delta_{izi}/\omega_{g}=0.052$, $\delta_{zii}/\omega_{g}=0.071$, $\delta_{zzi}/\omega_{g}=0.085$, and $\delta_{izz}/\omega_{g}=0.12$. Similar parameters are used for panels (f)--(g) in a mirror symmetric fashion with respect to the middle of the circuit.}
\label{fig:PhysNoise-XtalkCompToNum}
\end{figure*}

We find that the $ZZ$ crosstalk acts trivially in the control spectator case, since, based on Eqs.~(\ref{eq:PhysNoise-ContSpecXtalk Hg})--(\ref{eq:PhysNoise-ContSpecXtalk Hdelta}), the noise commutes with the ideal gate, i.e. $[H_g,H_{\delta}]=0$. Therefore, in the effective PL model, the only non-zero noise generator term is $\lambda_{zzi} = (\theta_{g}^2 \delta_{zzi}^2)/(4\omega_{g}^2)$ for $g = cx, \ cz$. On the other hand, a $ZZ$ crosstalk with the target spectator of a $CX_{\theta}$ gate, as shown in Fig.~\ref{fig:PhysNoise-ZZCrosstalkCases}(d), leads to a rather involved spreading of the crosstalk. In particular, rotating the noise term $IZZ$ by the $IXI$ and $ZXI$ terms in $H_g$ leads to effective non-zero PL generator terms $\lambda_{izz}$, $\lambda_{iyz}$, $\lambda_{zyz}$ and $\lambda_{zzz}$ (lower right column of Table~\ref{Tab:MagnusPert-3QZZCrosstalk}). This serves as the first example of weight-3 noise terms that are \textit{not} accounted for in the state-of-the-art sparse PL models. Our technique therefore can inform the learning methods what higher-weight PL noise terms are relevant to be added to the noise model.       

We further analyze $ZZ$ crosstalk between two adjacent two-qubit gates accounting for multiple cases described in the bottom row of Fig.~\ref{fig:PhysNoise-ZZCrosstalkCases}. We begin with the Lindblad Eq.~(\ref{eq:PhysNoise-CohNoise Lind}) with the noise Hamiltonian   
\begin{align}
H_{\delta} = \frac{\delta_{izzi}}{2}IZZI \;,
\label{eq:PhysNoise-4QXtalk Hdelta}
\end{align}
and the ideal Hamiltonian modeled as
\begin{align}
\begin{split}
 H_{g}  &= \frac{\omega_{cz}}{2} (IIII-IZII-ZIII+ZZII) \\
& +\frac{\omega_{cz}}{2} (IIII-IIIZ-IIZI+IIZZ) \;,
\end{split}
\label{eq:PhysNoise-4QXtalk H_CZxCZ}\\
H_{g}  & = \frac{\omega_{cx}}{2} (IXII-ZXII+IIIX-IIZX)\;, 
\label{eq:PhysNoise-4QXtalk H_CXxCX}\\
H_{g} & = \frac{\omega_{cx}}{2} (IXII-ZXII+IIXI-IIXZ) \;,
\label{eq:PhysNoise-4QXtalk H_CXxXC}
\end{align}
for implementing operations $CZ_{\theta}\otimes CZ_{\theta}$, $CX_{\theta}\otimes CX_{\theta}$, and $CX_{\theta}\otimes X_{\theta}C$, as in Figs.~\ref{fig:PhysNoise-ZZCrosstalkCases}(f)--\ref{fig:PhysNoise-ZZCrosstalkCases}(h), respectively. In case (h), as opposed to (g), the controlled gates operate in opposite directions such that the $ZZ$ noise acts between the neighboring target qubits. This case is practically relevant for CR based architectures as the controlled gates typically have well-defined native direction with a positive control-target detuning. 

Similar to the three-qubit analysis, $ZZ$ crosstalk noise acts trivially as it commutes when the ideal operation is $I_{\tau_g}$ or $CZ_{\theta} \otimes CZ_{\theta}$ shown in Figs.~\ref{fig:PhysNoise-ZZCrosstalkCases}(e)--\ref{fig:PhysNoise-ZZCrosstalkCases}(f). However, the interplay of $ZZ$ crosstalk and $CX_{\theta}$ operations is non-trivial. Our perturbative analysis reveals that the leading-order effective PL noise in the case of the $CX_{\theta}\otimes CX_{\theta}$ in Fig.~\ref{fig:PhysNoise-ZZCrosstalkCases}(g) is exactly the same as the three-qubit target spectator case of Fig.~\ref{fig:PhysNoise-ZZCrosstalkCases}(d) (compare the bottom right column of Table~\ref{Tab:MagnusPert-3QZZCrosstalk} to the left column of Table~\ref{Tab:MagnusPert-4QZZCrosstalk}). In the case of $CX_{\theta}\otimes X_{\theta}C$ operation in Fig.~\ref{fig:PhysNoise-ZZCrosstalkCases}(h), however, the noise does not commute with either of the gates, and therefore spreads the $IZZI$ crosstalk into 16 distinct weights-2, weight-3 and weight-4 effective PL generator terms (right column of Table~\ref{Tab:MagnusPert-4QZZCrosstalk}).  

\subsection{Comparison with numerical simulation}
\label{SubSec:NumComp}

To confirm the validity and demonstrate the utility of our Lindblad perturbation theory, we compare the results in Tables~\ref{Tab:MagnusPert-T1Decay}--\ref{Tab:MagnusPert-4QZZCrosstalk} to a direct numerical simulation of the noise. To this end, we first analyze two-qubit $CZ_{\pi/4}$ and $CX_{\pi/4}$ operations with simultaneously added $T_{1}$, $T_{2\phi}$ and coherent phase noise (results in Fig.~\ref{fig:PhysNoise-2QCompToNum}). Furthermore, having found that non-trivial $ZZ$ crosstalk occurs when the target qubits of $CX_{\theta}$ operations are involved, we also analyze the three-qubit case (d) and four-qubit case (h) of Fig.~\ref{fig:PhysNoise-ZZCrosstalkCases} (results in Fig.~\ref{fig:PhysNoise-XtalkCompToNum}). In all cases considered, adding up the perturbative $\lambda_k$ contributions for each individual coherent and incoherent noise mechanisms in Tables~\ref{Tab:MagnusPert-T1Decay}--\ref{Tab:MagnusPert-4QZZCrosstalk} leads to a precise breakdown of the overall numerical estimates for $\lambda_k$ parameters.      

In our numerical modeling, we simulate the following Lindblad equation using Qiskit Dynamics \cite{Puzzuoli_Qiskit_2023, Puzzuoli_Algorithms_2023}
\begin{align}
\begin{split}
\dot{\rho}(t) &= -i[H_g + H_{\delta}, \rho(t)] \\
&+ \sum_{j} \beta_{\downarrow j} \mathcal{D}[S_j^{-}]\rho(t) + \sum_{j} \frac{\beta_{\phi j}}{2} \mathcal{D}[Z_j^{-}]\rho(t) \;,
\end{split}
\label{eq:PhysNoise-FullNumLind}
\end{align}
with ideal Hamiltonian $H_g$, and noise Hamiltonian $H_{\delta} = H_{\text{phase}} + H_{\text{xtalk}}$ accounting for intra-gate phase noise and inter-gate crosstalk noise. Moreover, we consider $T_1$ and $T_{2\phi}$ noise acting on each individual qubit in the system. Simulating the Lindbladian in the interval $[0,\tau_g]$, we first compute the channel for the noisy operation $\tilde{\mathcal{G}} \equiv \exp(\mathcal{L}\tau_g)$ and the corresponding noise channel $\mathcal{N} \equiv \mathcal{U}_{g}^{-1} \tilde{\mathcal{G}}$. From this, we derive an effective PL model by a symplectic transformation of the diagonal elements $\vec{f}$ of the PTM for $\mathcal{N}$ as $\vec{\lambda} = -(1/2)S^{-1} \log(\vec{f})$, where $S$ is the symplectic matrix and $\vec{\lambda}$ are the PL noise generator parameters (see \cite{Berg_Probabilistic_2023} for details). We then compare numerical and perturbative PL generator parameters $\vec{\lambda}$ and fidelities $\vec{f}$.

Figure~\ref{fig:PhysNoise-2QCompToNum} summarizes our numerical versus perturbative analysis for the $CZ_{\pi/4}$ (top row) and $CX_{\pi/4}$ (bottom row) operations with weak continuous coherent and incoherent noise. In panels (d) and (i), for $CZ_{\pi/4}$ and $CX_{\pi/4}$, the overall numerically derived PL $\lambda_k$ parameters (blue bars) are well explained by the sum of corresponding $\lambda_k$ for each individual noise mechanism in Tables~\ref{Tab:MagnusPert-T1Decay}--\ref{Tab:MagnusPert-2QPhaseError}. The corresponding numerical and perturbative Pauli fidelities also show a good agreement in panels (e) and (j) . The numerical results confirm the role of commutativity of a given noise term with the ideal operation, where the $T_{2\phi}$ noise spreads over onto specific weight-2 PL parameters $\lambda_{k}$ only for $CX_{\pi/4}$ but not for $CZ_{\pi/4}$ (green bars). The $T_1$ noise, however, does not commute with either of the gates, so the noise spreads into weight-2 PL parameters in both cases (orange bars). Moreover, we find that the considered physical noise mechanisms result in an effective PL noise channel that is far from a uniform depolarizing noise model where, besides the non-uniform dependence on physical noise strengths, certain weight-2 PL parameters are zero ($\lambda_{xx}$, $\lambda_{xy}$, $\lambda_{yx}$ and $\lambda_{yy}$ for the $CZ_{\pi/4}$, and $\lambda_{xy}$, $\lambda_{xz}$, $\lambda_{yy}$, $\lambda_{yz}$, and $\lambda_{zx}$ for the $CX_{\pi/4}$).       

Figure~\ref{fig:PhysNoise-XtalkCompToNum} summarizes our crosstalk analysis for the $CX_{\pi/4}\otimes I$ (top two rows), and for the $CX_{\pi/4}\otimes X_{\pi/4}C$ (bottom row) operations with intra-gate phase noise and inter-gate $ZZ$ crosstalk as well as $T_1$ and $T_2$ noise on each qubit. First, we confirm the utility of our perturbative analysis as the sum over leading-order individual mechanisms explain the entire noise landscape in such rich multi-qubit circuits, as illustrated in panels (d) and (g) for the three- and four-qubit cases, respectively. We note again the non-uniform and sparse distribution of the noise generator $\lambda_k$ parameters where, with an approximate threshold of $10^{-5}$, there are only 17 (out of 63) and 36 (out of 255) non-zero generator terms for the two cases. Having the knowledge of non-zero generator terms can significantly simplify the PL learning step. Second, the numerical results confirm our perturbative finding that the $ZZ$ crosstalk can spread into higher-weight Pauli generators. For the three-qubit $CX_{\pi/4}\otimes I$ operation, the original $IZZ$ Hamiltonian term turns into non-negligible effective $IYZ$ and $ZYZ$, and slightly weaker $ZZZ$ generators. The crosstalk mixing for the four-qubit $CX_{\pi/4}\otimes X_{\pi/4}C$ operation is  much richer, consisting of 16 distinct weight-2, weight-3 and weight-4 rates found in Table~\ref{Tab:MagnusPert-4QZZCrosstalk}, which is corroborated by numerical analysis in panel (g). Importantly, the indices in which an effective Pauli generator due to $ZZ$ crosstalk emerge are unique (purple bars), such that there are no contributions from other considered physical sources. Our analysis provides a precise diagnostic pattern for each underlying noise mechanism, and can also inform the experiment what higher-weight PL should be included.
    

\section{Conclusion}

In this work, we introduce methods to efficiently compute the noise channel given a Lindblad description of a noisy multi-qubit operation. Our work serves as an intermediary step that connects Lindblad learning to QEM/QEC. We propose use cases both in the context of a generic numerical noise construction module (Fig.~\ref{fig:schematic_summary}(a), Sec.~\ref{SubSec:LindPert} and Fig.~\ref{fig:NumMagnusCXpiOv4}), and as an analytical tool to better understand the rich interplay of Lindblad noise mechanisms with entangling gate operations (Sec.~\ref{Sec:PhysNoise}, Appendix~\ref{App:WhyPauliLind}, Tables~\ref{Tab:MagnusPert-T1Decay}--\ref{Tab:MagnusPert-4QZZCrosstalk}, and Figs.~\ref{fig:PhysNoise-ZZCrosstalkCases}--\ref{fig:PhysNoise-XtalkCompToNum}). We provide a first-principles derivation of effective sparse PL models, which are commonly used in PEC and PEA. In particular, we present leading-order analytical expressions for the resulting sparse PL noise model due to various physically relevant Lindblad noise mechanisms, such as incoherent amplitude damping and pure dephasing, and coherent two-qubit phase noise and inter-gate crosstalk in three- and four-qubit circuit scenarios. 

Our analytical characterization of the sparse PL noise model due to physical noise provides valuable insights about the PL model's origin and noise structure. First, we observe that the sparse PL model generator has a rich dependence on the underlying physical noise mechanisms, where a given $\lambda_k$ generator term could originate from multiple physical sources. Importantly, for weak noise, our leading-order perturbative $\lambda_k$ expressions due to each mechanism provide a precise breakdown of the overall $\lambda_k$ simulated with all noise sources acting simultaneously (Figs.~\ref{fig:PhysNoise-2QCompToNum}--\ref{fig:PhysNoise-XtalkCompToNum}). Second, we find that depending on the commutativity of the noise with the ideal gate, and the entangling nature of the ideal gate, local (weight-1) incoherent noise such as $T_1$ and $T_{2\phi}$ dissipators can spread into effective non-local (weight-2) PL generator terms. More drastically, we show analytically and confirm with numerical simulations that weight-2 coherent $ZZ$ crosstalk can turn into effective weight-3 and weight-4 PL $\lambda_k$ terms. Third, our analysis provides practical signatures for identifying the underlying physical noise mechanism based on the observed behavior of the measured PL model. In particular, PL $\lambda_k$ parameters have a distinct dependence on the gate angle, which is linear and quadratic for coherent and incoherent noise sources, respectively. A sinusoidal pairwise spreading of the PL $\lambda_k$ terms indicates a physical noise that does not commute with the gate (Appendix~\ref{App:AngleDepofPLGen}). Moreover, observing certain patterns of high-weight non-zero PL terms can indicate a spreading of crosstalk.  
    
Looking ahead, our work paves the way for a lower-level physical understanding of operational noise models. One important direction is the development of more realistic sparse PL noise models for future PEC and PEA applications \cite{Temme_Error_2017, Endo_practical_2018, Kandala_Error_2019, Berg_Probabilistic_2023, Kim_Evidence_2023}, with targeted higher-weight generator terms informed by our noise spreading findings. Another relevant application of our perturbative analysis is to understand the fundamental symmetries and learnable degrees of freedom \cite{Chen_Learnability_2023, Seif_Entanglement_2024, Chen_Efficient_2024} from a physical perspective. Despite our focus on the PL noise model, adopted in PEC and PEA for twirled noise, we emphasize that the scope of our noise construction method is broader. In particular, we expect our method to have immediate applications in improved noise modeling of QEC \cite{Schwartzman_Modeling_2024}, and also for noise-aware codes and decoders \cite{Leung_Approximate_1997, Fletcher_Optimum_2007, Nickerson_Analysing_2019, Schwartzman_Modeling_2024}.   


\section{Acknowledgements}
We acknowledge helpful discussions with IBM quantum team members Ken Xuan Wei, Bradley Mitchell, Christopher Wood, Abhinav Kandala, David Layden, Emily Pritchett, and Andrew Cross.

\appendix
\section{Time-independent Lindblad model}
\label{App:LindModel}

Under the assumption that the noise is Markovian, its generator can be parametrized as a time-independent Lindbladian \cite{Wolf_Dividing_2008, Wolf_Assessing_2008}, which maps the end-to-end action of a given quantum operation. To this end, we adopt the following Lindblad model:
\begin{align}
& \dot{\rho}(t) = - i [\sum\limits_j \alpha_j P_j,\rho(t)] + \sum\limits_{jk} \beta_{jk} \mathcal{D}[P_j,P_k]\rho(t)  \;,
\label{Eq:LindModel-lindblad 1} \\
& \mathcal{D}[P_j,P_k]\rho(t) \equiv P_j \rho(t) P_k^{\dag} - \frac{1}{2} \{P_k^{\dag} P_j,\rho(t) \}\;,
\label{Eq:LindModel-Def of D[Pi,Pj]} 
\end{align}
where the overall Hamiltonian is parametrized as $H = \sum_{j} \alpha_j P_j$, with real coefficients $\alpha_j$, and $\beta_{jk}$ describes the damping (incoherent) processes, where $\bm{\beta}$ is a positive semi-definite matrix. 

Furthermore, in our noise synthesis, we can separate the overall Hamiltonian $H\equiv H_g + H_{\delta}$ as an ideal $H_g$, for which the ideal gate unitary is $U_g \equiv \exp(-iH_g\tau_g)$ at gate time $\tau_g$, and a Hamiltonian (coherent) error contribution $H_{\delta}$. Some physically motivated coherent noise mechanisms are single-qubit $X$ (bit flip) and $Z$ (phase flip), over-rotation, and coherent $ZZ$ both within a two-qubit gate or with adjacent spectator qubits. Another form of coherent error is leakage. However, by the adoption of a two-level model in Eqs.~(\ref{Eq:LindModel-lindblad 1})--(\ref{Eq:LindModel-Def of D[Pi,Pj]}) leakage is not accounted for from the outset.      

To illustrate the generic off-diagonal form of $\bm{\beta}$ in Eqs.~(\ref{Eq:LindModel-lindblad 1})--(\ref{Eq:LindModel-Def of D[Pi,Pj]}), consider the amplitude damping ($T_1$) Lindblad generator $\beta_{\downarrow} (S^{-}\bullet S^{+}-(1/2)\{S^{+}S^{-},\bullet\})$. Expressing the raising/lowering operator in terms of Pauli operators as $S^{-}=(X+iY)/2$ and $S^{+}=(X-iY)/2$, we can re-write the amplitude damping Lindblad equation in the canonical form~(\ref{Eq:LindModel-Def of D[Pi,Pj]}) as:
\begin{align}
\begin{split}
\dot{\rho} & =  \frac{\beta_{\downarrow}}{4} \big(X \rho X + Y \rho Y -i X \rho Y \\
& + i Y \rho X  -Z \rho I - I \rho Z - 2 \rho \big) \\
& = \frac{\beta_{\downarrow}}{4}\left[X \rho X - \rho\right] + \frac{\beta_{\downarrow}}{4}\left[Y \rho Y - \rho\right] \\
&+ \frac{-i\beta_{\downarrow}}{4}\left[X \rho Y - (1/2) \{YX,\rho\}\right] \\
&+ \frac{i\beta_{\downarrow}}{4}\left[Y \rho X - (1/2) \{XY,\rho\}\right] \;.	
\end{split}
\label{Eq:LindModel-Pauli rep of AD Lindblad}
\end{align} 
Therefore, the Pauli representation of the amplitude damping dissipator matrix is found as:
\begin{align}
\bm{\beta}_{\text{AD}} = \frac{\beta_{\downarrow}}{4} \begin{bmatrix}
0 & 0 & 0 & 0 \\
0 & 1 & -i & 0 \\
0 & +i & 1 & 0 \\
0 & 0 & 0 & 0
\end{bmatrix} \;.
\label{Eq:LindModel-Pauli rep of AD Lindblad}
\end{align}
Another physically relevant incoherent noise mechanism is pure dephasing, whose Lindblad generator $(\beta_{\phi}/2) (Z\bullet Z-\bullet)$ has a simple diagonal representation   
\begin{align}
\bm{\beta}_{\text{PD}} = \frac{\beta_{\phi}}{2} \begin{bmatrix}
0 & 0 & 0 & 0 \\
0 & 0 & 0 & 0 \\
0 & 0 & 0 & 0 \\
0 & 0 & 0 & 1
\end{bmatrix} \;.
\end{align}

Starting from the Lindblad model described here, we explore the interplay of relevant coherent and incoherent noise mechanisms with two-qubit gates in multi-qubit circuit scenarios in Appendix~\ref{App:WhyPauliLind} and Sec.~\ref{Sec:PhysNoise}.

\section{Noise extraction via interaction-frame representation}
\label{App:IntFrameRep}
\begin{figure*}[t!]
\centering
\includegraphics[scale=0.74]{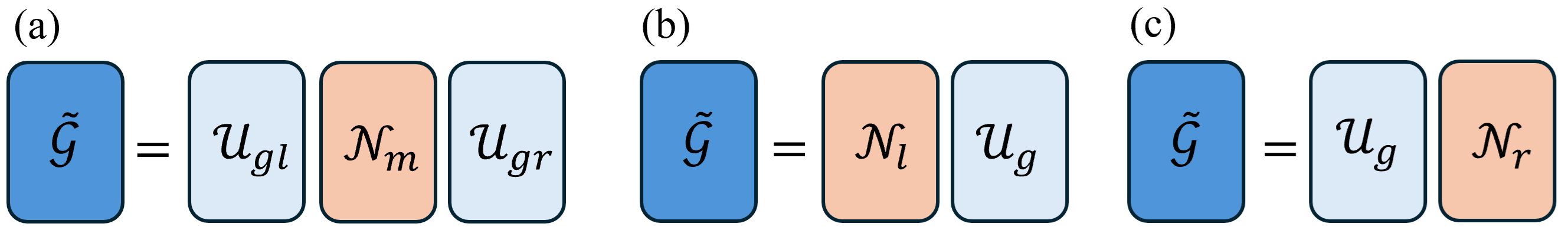}
\caption{\textbf{Various noise decompositions via interaction frame representation} - (a) generalized decomposition using the time displaced frame as given in Eq.~(\ref{Eq:IntFrameRep-int rep 2}) with the noise acting in the middle, (b) standard (left) decomposition with the noise acting first corresponding to the time shift choice $\tau=0$, (c) unconventional (right) decomposition with the ideal gate acting first corresponding to the time shift choice $\tau=\tau_g$. In the schematics, $\tilde{\mathcal{G}} \equiv \exp(\mathcal{L}\tau_g)$ is the learned noisy operation, $\mathcal{U}_g\equiv \exp(\mathcal{H}_g\tau_g)$ is the ideal gate, and $\mathcal{N}_m \equiv \mathcal{T} \exp[\int_{0}^{\tau_g} dt' \mathcal{L}_I(t';\tau_s)]$ is the decomposition with noise in the middle via a generic time shift.}
\label{Fig:IntFrameRep-NoiseDecompSchematics}
\end{figure*}

To separate the noise from the ideal operation, it is natural to describe the time evolution in the interaction frame with respect to the gate Hamiltonian $H_g$. Expressing the density matrix $\rho(t)$ in the interaction frame
\begin{align}
\rho(t) \equiv e^{-i H_g t} \rho_I(t) e^{+i H_g t} \;,
\label{Eq:IntFrameRep-def of tilde(rho)}
\end{align}
and replacing in the starting Lindblad Eq.~(\ref{Eq:LindModel-lindblad 1}), we find 
\begin{align}
\begin{split}
e^{-i H_g t}\dot{\rho}_I(t)e^{+i H_g t} = -i[\sum\limits_j \delta_j P_j(t),e^{-i H_g t}\rho_I(t)e^{+i H_g t}] \\
+ \sum\limits_{jk} \beta_{jk}\Big(P_j e^{-i H_g t}\rho_I(t)e^{+i H_g t} P_k^{\dag} \\
- \frac{1}{2}\{P_k^{\dag}P_j,e^{-i H_g t}\rho_I(t)e^{+i H_g t}\}\Big)  \;.
\end{split}
\label{Eq:IntFrameRep-lindblad 2}
\end{align}
Multiplying Eq.~(\ref{Eq:IntFrameRep-lindblad 2}) from the left and right by $\exp(+iH_g t)$ and $\exp(-iH_g t)$, respectively, we obtain the interaction frame representation as: 
\begin{align}
\begin{split}
& \dot{\rho}_I(t) = \mathcal{L}_I[\rho_I(t)] \equiv  -i[\sum\limits_j \delta_j P_{jI}(t),\rho_I(t)]  \\
&+ \sum\limits_{jk} \beta_{jk}\left(P_{jI}(t) \rho_I(t) P_{kI}^{\dag}(t) - \frac{1}{2}\{P_{kI}^{\dag}(t)P_{jI}(t),\rho_I(t)\}\right)
\end{split}
\label{Eq:IntFrameRep-lindblad 3}
\end{align}
with corresponding rotated Pauli operators defined as
\begin{align}
P_{jI}(t) \equiv e^{+i H_g t} P_{j} e^{-i H_g t} \;.
\label{Eq:IntFrameRep-Def of tilde(P)_i}
\end{align}
Consequently, the overall time evolution in the interval $[0,\tau_g]$ takes a product form as: 
\begin{align}
e^{\mathcal{L} \tau_g} = \mathcal{U}_g(\tau_g) \mathcal{T} e^{\int_{0}^{\tau_g} dt'\mathcal{L}_I(t')} \;,
\label{Eq:IntFrameRep-int rep 1}
\end{align}
where $\mathcal{U}_g(t)\equiv \exp(-i\mathcal{H}_g t)$ is the ideal Hamiltonian evolution, with $\mathcal{H}_g \equiv [H_g,\bullet]$ as the Hamiltonian superoperator, and $\mathcal{T}$ is the time-ordering operator given the time-dependent Lindbladian $\mathcal{L}_I(t)$.

In decomposition~(\ref{Eq:IntFrameRep-int rep 1}), the noise acts before the ideal operation. For general purpose noise synthesis, it is helpful to have the flexibility to partition the noise arbitrarily before or after the ideal operation. With this aim, we introduce a generalized interaction-frame representation
\begin{align}
\rho(t) \equiv e^{-i H_g (t-\tau_s)} \rho_I(t;\tau_s) e^{+i H_g (t-\tau_s)} \;,
\label{Eq:IntFrameRep-def of tilde(rho(t;tau))}
\end{align}
where $\tau_s$ is an arbitrary time displacement. Going through a similar derivation described above, we find a modified decomposition
\begin{align}
e^{\mathcal{L} \tau_g} = \mathcal{U}_g(\tau_g-\tau_s) \mathcal{T} e^{\int_{0}^{\tau_g} dt' \mathcal{L}_I(t';\tau_s)} \mathcal{U}_g(\tau_s) \;,
\label{Eq:IntFrameRep-int rep 2}
\end{align}
where now fractions of the ideal operation wrap the noise from both sides. Choosing $\tau_s=0$ results in the standard decomposition~(\ref{Eq:IntFrameRep-int rep 1}), while setting $\tau_s=\tau_g$ the order is reversed where the noise acts after the ideal operation, resulting in the following noise definitions:    
\begin{subequations}
\begin{align}
& \mathcal{N}_m \equiv \mathcal{T} e^{\int_{0}^{\tau_g} dt' \mathcal{L}_I(t';\tau_s)} \;,
\label{Eq:IntFrameRep-Def of Nm}\\
& \mathcal{N}_l \equiv \mathcal{T} e^{\int_{0}^{\tau_g} dt' \mathcal{L}_I(t';\tau_s=0)} \;,
\label{Eq:IntFrameRep-Def of Nl}\\
& \mathcal{N}_r \equiv \mathcal{T} e^{\int_{0}^{\tau_g} dt' \mathcal{L}_I(t';\tau_s=\tau_g)} \;.
\label{Eq:IntFrameRep-Def of Nr}  
\end{align}   
\end{subequations}
Figure~(\ref{Fig:IntFrameRep-NoiseDecompSchematics}) visualizes various decompositions of a noisy operation in terms of ideal and noise components.

Lastly, we note that $\mathcal{L}_I$ in Eq.~(\ref{Eq:IntFrameRep-lindblad 3}) treats Hamiltonian (coherent) and dissipative (incoherent) components on equal footing. In an alternative decomposition, we can implement two successive transformations: (i) express $\mathcal{L}$ in the interaction frame with respect to $H_g + H_{\delta}$, (ii) express $H_{\delta}$ in the interaction frame with respect $H_g$. Such a decomposition separates the ideal  Hamiltonian and coherent and incoherent errors into a product form.


\section{Time-dependent Lindblad perturbation theory}
\label{App:TDLindPT}

The interaction-frame Lindbladian is time dependent, hence the time-ordering operator cannot be computed exactly except for special cases. To compute the noise, we therefore employ time-dependent Magnus and Dyson perturbation theories. Such an expansion is practical assuming that both the coherent and incoherent noise strengths are sufficiently weaker than the intended coherent gate interaction, i.e. $\delta_{j}/\omega_{g} \ll 1$ and $\beta_{ij}/\omega_{g} \ll 1$ with $\omega_g$ as the dominant Hamiltonian coefficient.

In Magnus perturbation theory, we solve for an effective generator of the dynamics. For our problem of computing the noise, we express the time evolution operator in the interaction frame as  
\begin{align}
\mathcal{T} e^{\int_{0}^{t} dt'\mathcal{L}_I(t')} = e^{\Omega(t,0)} \;,
\label{Eq:Def of Magnus G}
\end{align}
with the generator $\Omega(t,0)$ to be solved for perturbatively. Here, for notation simplicity, we have dropped the time translation argument $\tau_s$. Up to the fourth order in $\mathcal{L}_I$, the Magnus expansion reads \cite{Magnus_Exponential_1954, Blanes_Magnus_2009, Blanes_Pedagogical_2010, Puzzuoli_Algorithms_2023}:
\begin{align}
\Omega_1(t,0)   &= \int_{0}^{t} dt'\mathcal{L}_I(t') \;,
\label{Eq:TDLindPT-G1 Sol} \\
\Omega_2(t,0)   &= \frac{1}{2} \int_{0}^{t} dt' \int_{0}^{t'} dt'' [\mathcal{L}_I(t'),\mathcal{L}_I(t'')] \;,
\label{Eq:TDLindPT-G2 Sol} \\
\begin{split}
\Omega_3(t,0)   &= \frac{1}{6} \int_{0}^{t} dt' \int_{0}^{t'} dt''\int_{0}^{t''} dt''' \\
                &\Big( [\mathcal{L}_I(t'),[\mathcal{L}_I(t''),\mathcal{L}_I(t''')]]  \\
                &+ [\mathcal{L}_I(t'''),[\mathcal{L}_I(t''),\mathcal{L}_I(t')]]\Big) \;,
\label{Eq:TDLindPT-G3 Sol}
\end{split} \\
\begin{split}
\Omega_4(t,0)   &= \frac{1}{12} \int_{0}^{t} dt' \int_{0}^{t'} dt''\int_{0}^{t''} dt'''\int_{0}^{t'''} dt'''' \\
                &\Big([\mathcal{L}_I(t'),[\mathcal{L}_I(t''),[\mathcal{L}_I(t'''),\mathcal{L}_I(t'''')]]]  \\
                &+ [\mathcal{L}_I(t'),[[\mathcal{L}_I(t''),\mathcal{L}_I(t''')],\mathcal{L}_I(t'''')]] \\
                &+ [[[\mathcal{L}_I(t'),\mathcal{L}_I(t'')],\mathcal{L}_I(t''')],\mathcal{L}_I(t'''')] \\
                &+ [\mathcal{L}_I(t''),[\mathcal{L}_I(t'''),[\mathcal{L}_I(t''''),\mathcal{L}_I(t')]]]\Big) \;,
\label{Eq:TDLindPT-G4 Sol}
\end{split}
\end{align}
To compute the noise channel under Magnus perturbation, we exponentiate an approximate (truncated) effective generator according to Eqs.~(\ref{Eq:Def of Magnus G})--(\ref{Eq:TDLindPT-G4 Sol}).

For symbolic calculation of the noise, computing the matrix exponential in Eq.~(\ref{Eq:Def of Magnus G}) is only possible for simple problems. Under the Dyson perturbation theory, we perform an additional Taylor expansion \cite{Burum_Magnus_1981, Salzman_Alternative_1985, Malekakhlagh_Mitigating_2022, Puzzuoli_Algorithms_2023} to write:
\begin{align}
\begin{split}
\mathcal{T} e^{\int_{0}^{t} dt' \mathcal{L}_I(t')} &= \mathcal{I}+\Omega_1+\Omega_2+\frac{1}{2}\Omega_1^2 \\
&+\Omega_3 + \frac{1}{2}(\Omega_1\Omega_2 + \Omega_2\Omega_1) + \frac{1}{6}\Omega_1^3\\
&+\Omega_4+\frac{1}{2}(\Omega_1\Omega_3+\Omega_3\Omega_1)+\Omega_2^2\\
&+\frac{1}{6} (\Omega_1^2\Omega_2+\Omega_1\Omega_2\Omega_1+\Omega_2\Omega_1^2)\\
&+\frac{1}{24}\Omega_1^4+\mathcal{O}(\mathcal{L}_I^5)
\end{split}
\label{Eq:TDLindPT-N_Ch Dyson Sol}
\end{align} 
where we have used the simplified notation $\Omega_j\equiv \Omega_j(t,0)$. The results in Fig.~\ref{fig:NumMagnusCXpiOv4} are computed using the fourth-order Magnus and Dyson expressions provided in this Appendix.   


\section{Emergence of effective PL model for twirled noise}
\label{App:WhyPauliLind}

The sparse PL model was introduced to capture the twirled noise of CR based quantum processors with sparse connectivity \cite{Berg_Probabilistic_2023}. The ansatz for the twirled noise is taken as $\mathcal{N}_{\text{PL}} \equiv \exp[\sum_k \lambda_k (P_k \bullet P_k^{\dag} - I \bullet I)]$, with $\lambda_k$ as the model coefficients, which are inferred from the learned/measured Pauli fidelities. In this Appendix, we provide further motivation for such a model, and show that twirling a perturbatively derived Lindblad noise channel results in the PL noise form. We analyze multiple physically relevant examples of single- and two-qubit noisy gates for which we provide perturbative expressions for the resulting PL model coefficients.   

\subsection{Noise shaping via Pauli twirl}
\label{SubApp:Twirl}

Twirling a quantum channel amounts to averaging the channel over a group. Here, we study the application of Pauli twirling on the Lindblad noise channel as defined in Appendix~\ref{App:IntFrameRep}. Twirling a channel $\mathcal{N}$ over the Pauli group is defined as 
\begin{align}
\mathcal{N}_{\text{PL}} (\rho) \equiv  \frac{1}{|P|}\sum\limits_{k} P_k^{\dag}\mathcal{N}(P_k \rho P_k^{\dag})P_k \;.
\label{Eq:WhyPauliLind-Def of twirl}
\end{align}
Using the superoperator notation, such that $A \bullet B \leftrightarrow A \otimes B^T$, the action of Pauli twirl is clearer as an average over individual (Pauli) rotations:
\begin{align}
\begin{split}
\mathcal{N}_{\text{PL}} & = \frac{1}{|P|}\sum\limits_{k} (P_k^{\dag} \otimes P_k^{T})\mathcal{T} e^{\int_{0}^{\tau_g} dt' \mathcal{L}_I(t')}  (P_k \otimes P_k^{*}) \\
& = \frac{1}{|P|}\sum\limits_{k} \mathcal{T} e^{\int_{0}^{\tau_g} dt' (P_k^{\dag} \otimes P_k^{T}) \mathcal{L}_I(t')(P_k \otimes P_k^{*})} \;,
\end{split}
\label{Eq:WhyPauliLind-mathcal(E)_twirl}
\end{align}
where in the second line we used the fact that each term is a unitary transformation at the superoperator level so that it could apply directly on the generator (Lindbladian) of the channel. Furthermore, $\mathcal{L}_I(t')$ is the superoperator representation of Eq.~(\ref{Eq:IntFrameRep-lindblad 3}) as
\begin{align}
\begin{split}
\mathcal{L}_I(t')  = &-i\sum\limits_j \delta_j \Big[P_{jI}(t') \otimes I - I \otimes P_{jI}^T(t') \Big]\\
& +\sum\limits_{ij} \beta_{jk}\Big[P_{jI}(t') \otimes P_{kI}^{*}(t') \\
& - \frac{1}{2} P_{kI}^{\dag}(t')P_{jI}(t')\otimes I - \frac{1}{2} I \otimes P_{jI}^{T}(t')P_{kI}^{*}(t') \Big] \;.
\label{Eq:WhyPauliLind-SupOp Rep of mathcal(L)}
\end{split}
\end{align}
To assess whether the noise can be generated via a PL noise model, and characterize the generator $\lambda_k$ parameters, we take the matrix logarithm of Eq.~(\ref{Eq:WhyPauliLind-mathcal(E)_twirl}) as $\mathcal{L}_{\text{PL}} \equiv \log (\mathcal{N}_{\text{PL}})$. Based on the second line of Eq.~(\ref{Eq:WhyPauliLind-mathcal(E)_twirl}), this operation can be viewed as an operator extension of the log-sum-exp (LSE) function \cite{Nielsen_Guaranteed_2016} of the individual twirl terms. 

In the following, we apply the Magnus perturbation and twirl to numerous physically motivated noise mechanisms such as amplitude damping, pure dephasing, coherent phase ($Z$ and $ZZ$) and inter-gate $ZZ$ crosstalk. We consider cases covering single-qubit, isolated two-qubit $CZ_{\theta}$ and $CX_{\theta}$ gates, as well as various three-qubit and four-qubit crosstalk scenarios. We note that the single-qubit problems analyzed in Appendices~\ref{SubApp:Id+AD+PD}--\ref{SubApp:X_th+AD+PD} serve as idealized examples, where single-qubit twirling operations are considered ideal in theory for instructive purposes, whereas in practice they are implemented using noisy single-qubit gates. We summarize the results of our perturbation for the PL parameters in Tables~\ref{Tab:MagnusPert-T1Decay}--\ref{Tab:MagnusPert-4QZZCrosstalk}, organized in terms of the underlying physical noise process, for amplitude damping, pure dephasing, two-qubit coherent phase error, three-qubit $ZZ$ crosstalk, and four-qubit $ZZ$ crosstalk, respectively. \\     

\subsection{Single-qubit identity gate with amplitude damping and pure dephasing}
\label{SubApp:Id+AD+PD}

We begin with the simple example of an identity operation under amplitude damping and pure dephasing noise on a single qubit. We show that the twirled channel can be represented in terms of a PL generator. This problem is simple enough such that exact solutions of the model parameters exist without resorting to perturbation. 

Consider the Lindbladian for amplitude damping as in Eq.~(\ref{Eq:LindModel-Pauli rep of AD Lindblad}):
\begin{align}
\begin{split}
\mathcal{L}_{\text{AD}} & = \frac{\beta_{\downarrow}}{4} \Big(X \otimes X^{T} + Y \otimes Y^{T} -i X \otimes Y^{T} \\
&+ i Y \otimes X^{T} -Z \otimes I^{T} - I \otimes Z^{T} - 2 I \otimes I^{T} \Big) \;.
\end{split}
\label{Eq::WhyPauliLind-Pauli rep of AD Lindblad}
\end{align}
Application of the $I$ and $Z$ terms in the twirl leave this map invariant:
\begin{align}
\begin{split}
(I^{\dag} \otimes I^{T}) \mathcal{L}_{\text{AD}} (I \otimes I^{*}) & = (Z^{\dag} \otimes Z^{T}) \mathcal{L}_{\text{AD}} (Z \otimes Z^{*}) \\
& =\mathcal{L}_{\text{AD}} = \mathcal{L}_{\text{diag}} + \mathcal{L}_{\text{off-diag}}
\end{split}
\label{Eq::WhyPauliLind-L/Z twirl} \;,
\end{align}
where in the last step we have re-expressed $\mathcal{L}_{\text{AD}}$ in terms of its diagonal $\mathcal{L}_{\text{diag}}\equiv (X \otimes X^{T} + Y \otimes Y^{T} - 2 I \otimes I^{T})$ and off-diagonal $\mathcal{L}_{\text{off-diag}}\equiv (-i X \otimes Y^{T} + i Y \otimes X^{T} -  Z \otimes I^{T}-  I \otimes Z^{T})$ parts. In contrast, $X$ and $Y$ twirl terms change the sign of the off-diagonal parts:

\begin{align}
\begin{split}
& (X^{\dag} \otimes X^{T}) \mathcal{L}_{\text{AD}} (X \otimes X^{*}) = (Y^{\dag} \otimes Y^{T}) \mathcal{L}_{\text{AD}} (Y \otimes Y^{*})
\\
&= \frac{\beta_{\downarrow}}{4} \Big(X \otimes X^{T} + Y \otimes Y^{T} +i X \otimes Y^{T} - i Y \otimes X^{T} \\
&+Z \otimes I^{T} + I \otimes Z^{T} - 2 I \otimes I^{T} \Big) = \mathcal{L}_{\text{diag}}-\mathcal{L}_{\text{off-diag}} \;.
\end{split}
\label{Eq:WhyPauliLind-X/Y twirl}
\end{align}

Therefore, based on the second line of Eq.~(\ref{Eq:WhyPauliLind-mathcal(E)_twirl}), the twirled amplitude damping channel reads:
\begin{align}
\begin{split}
\mathcal{N}_{\text{PL}} = \frac{1}{2}\Big[\exp{(\mathcal{L}_{\text{diag}}\tau_g+\mathcal{L}_{\text{off-diag}}\tau_g)} \\
+ \exp{(\mathcal{L}_{\text{diag}}\tau_g-\mathcal{L}_{\text{off-diag}}\tau_g)}\Big] \;.
\end{split}
\label{Eq:WhyPauliLind-mathcal(L)_(AD,twirled)}
\end{align}
In general, the diagonal and off-diagonal parts of the twirled Lindbladian do not commute, hence the matrix exponential of each term should be computed independently. The resulting twirled AD channel reads:
\begin{align}
\begin{split}
\mathcal{N}_{\text{PL}} & = \frac{1}{4}(1+e^{-\beta_{\downarrow}\tau_g}+e^{-\beta_{\downarrow}\tau_g/2})(I\otimes I^{T}) \\
&+ \frac{1}{4}(1+e^{-\beta_{\downarrow}\tau_g}-e^{-\beta_{\downarrow}\tau_g/2})(Z\otimes Z^{T})\\
& + \frac{1}{4}(1-e^{-\beta_{\downarrow}\tau_g})(X\otimes X^{T} + Y\otimes Y^{T}) \;.
\end{split}
\label{Eq:WhyPauliLind-N_(AD,twirled)}
\end{align}
We conclude that twirling the amplitude damping channel results in a diagonal $\chi$-representation \cite{Wood_Tensor_2011} as in Eq.~(\ref{Eq:WhyPauliLind-N_(AD,twirled)}). We then ask whether Eq.~(\ref{Eq:WhyPauliLind-N_(AD,twirled)}) can be generated from a PL generator of the form $\mathcal{L}_{\text{PL}} \equiv \sum_{k=x,y,z} \lambda_k (P_k\otimes P_k^{T} - I\otimes I)$ such that $\mathcal{N}_{\text{PL}} = \exp(\mathcal{L}_{\text{PL}})$. Taking the matrix logarithm, we find that $\mathcal{L}_{\text{PL}} = \mathcal{L}_{\text{diag}}\tau_g$ where $\lambda_x=\lambda_y=(\beta_{\downarrow}\tau_g)/4$ and $\lambda_z=0$.

Next, the case of an identity operation with pure dephasing $\mathcal{L}_{\text{PD}} \equiv (\beta_{\phi}/2) (Z \otimes Z^T - I \otimes I)$ is trivial given that $\mathcal{L}_{\text{PD}}$ remains invariant under each Pauli twirl transformation. Therefore, the noise channel is found as $\mathcal{N}_{\text{PL}}=\exp(\mathcal{L}_{\text{PL}})$ where $\mathcal{L}_{\text{PL}} = \mathcal{L}_{\text{PD}}\tau_g$ corresponds to model coefficients $\lambda_x = \lambda_y =0$ and $\lambda_z = (\beta_{\phi}\tau_g)/2$. 

Lastly, for the case of simultaneous amplitude damping and pure dephasing we can follow a similar derivation as in Eqs.~(\ref{Eq::WhyPauliLind-L/Z twirl})--(\ref{Eq:WhyPauliLind-N_(AD,twirled)}), where the overall twirled noise reads: 
\begin{align}
\begin{split}
\mathcal{N}_{\text{PL}} & = \frac{1}{4}(1+e^{-\beta_{\downarrow}\tau_g}+e^{-(\beta_{\downarrow}/2+\beta_{\phi})\tau_g})(I\otimes I^{T}) \\
&+ \frac{1}{4}(1+e^{-\beta_{\downarrow}\tau_g}-e^{-(\beta_{\downarrow}/2+\beta_{\phi})\tau_g})(Z\otimes Z^{T})\\
& + \frac{1}{4}(1-e^{-\beta_{\downarrow}\tau_g})(X\otimes X^{T} + Y\otimes Y^{T}) \;,
\end{split}
\label{Eq:WhyPauliLind-N_(AD+PD,twirled)}
\end{align}
which can be generated as a sum of the effective AD and PD PL generators with coefficients $\lambda_x = \lambda_y =(\beta_{\downarrow}\tau_g)/4$ and $\lambda_z = (\beta_{\phi}\tau_g)/2$.

\subsection{Single-qubit $X_{\theta}$ gate with amplitude damping and pure dephasing}
\label{SubApp:X_th+AD+PD}

We next consider the case of a single-qubit arbitrary-angle $X_{\theta}\equiv \exp[-i(\theta/2)X]$ gate with amplitude damping and pure dephasing, obeying the Lindblad equation:  
\begin{align}
\begin{split}
\dot{\rho}(t) = - i \left[\frac{\omega_x}{2}X,\rho(t)\right] + \beta_{\downarrow} \mathcal{D}[S^{-}]\rho(t) + \frac{\beta_{\phi}}{2} \mathcal{D}[Z]\rho(t) \;.
\end{split}
\label{Eq:WhyPauliLind-1Q Hz+SX+GenDiss Lind}
\end{align}
In the following, we compute the effective twirled noise channel, and show that it can be re-cast into a PL form.

Moving to the interaction frame with respect to $H_g \equiv (\omega_x/2) X$, integrating the Lindbladian in the interval $[0,\tau_g = \theta/\omega_x]$ up to the leading order in the Dyson series, and applying single-qubit twirl results in the following noise channel:
\begin{align}
\begin{split}
\mathcal{N}_{\text{PL}} & \equiv \frac{1}{|P|}\sum\limits_{k} \mathcal{T} e^{\int_{0}^{\tau_g} dt' (P_k^{\dag} \otimes P_k^{T}) \mathcal{L}_I(t')(P_k \otimes P_k^{*})} \\
& =\begin{bmatrix}
w_{1} & 0 & 0 & w_2 \\ 
0 & w_{3} & w_{4} & 0 \\ 
0 & w_{4} & w_{3} & 0 \\ 
w_{2} & 0 & 0 & w_1 
\end{bmatrix} + O(\beta^2)\;,
\end{split}
\label{Eq:WhyPauliLind-1Q SX+PD mathcal(E)_twirl}
\end{align}
with weights $w_i$ found up to to the leading (linear) order in $\beta_{\downarrow}/\omega_x$ and $\beta_{\phi}/\omega_x$ as
\begin{subequations}
\begin{align}
& w_1 = 1 + \frac{\sin \left(2 \theta\right) \left(2 \beta _{\phi}-\beta_{\downarrow}\right)-2
   \theta \left(3\beta_{\downarrow}+2\beta_{\phi}\right)}{16
   \omega _x} \;,
\label{Eq:WhyPauliLind-1Q SX+GenDiss expanded w1}
\end{align}
\begin{align}
& w_2 = - \frac{\sin \left(2 \theta\right) \left(2 \beta_{\phi}-\beta_{\downarrow}\right)-2
   \theta \left(3\beta_{\downarrow}+2\beta_{\phi}\right)}{16
   \omega _x} \;,
\label{Eq:WhyPauliLind-1Q SX+GenDiss expanded w2}
\end{align}
\begin{align}
& w_3 = 1+ \frac{\sin \left(2 \theta\right) \left(\beta _{\downarrow}-2\beta _{\phi}\right)-2
   \theta \left(5 \beta_{\downarrow}+6 \beta_{\phi}\right)}{16 \omega _x}\;,
\label{Eq:WhyPauliLind-1Q SX+GenDiss expanded w3}
\end{align}
\begin{align}
& w_4 = \frac{(2\beta_{\phi}-\beta_{\downarrow})[\sin(2\theta)-2\theta ]}{16
   \omega_x} \;.
\label{Eq:WhyPauliLind-1Q SX+GenDiss expanded w4}
\end{align}
\end{subequations}
Trace preservation of a channel of the form~(\ref{Eq:WhyPauliLind-1Q SX+PD mathcal(E)_twirl}) requires $w_1+w_2=1$ which holds based on the explicit expressions~(\ref{Eq:WhyPauliLind-1Q SX+GenDiss expanded w1})--(\ref{Eq:WhyPauliLind-1Q SX+GenDiss expanded w2}).

Taking the matrix logarithm of Eq.~(\ref{Eq:WhyPauliLind-1Q SX+PD mathcal(E)_twirl}), the noise channel can be generated via a modified PL model $\Omega_{\text{PL}} \equiv \sum_k \lambda_k (P_k \otimes P_k^{T}-I \otimes I^{T})$ with parameters $\lambda_k$ up to the leading order found as:
\begin{subequations}
\begin{align}
& \lambda_x = \frac{\theta}{4}\frac{\beta_{\downarrow}}{\omega_x} \;,
\label{Eq:WhyPauliLind-1Q SX+GenDiss lambda_x}\\
& \lambda_y = \frac{2\theta + \sin(2\theta)}{16}\frac{\beta_{\downarrow}}{\omega_x} + \frac{2\theta - \sin(2\theta)}{8}\frac{\beta_{\phi}}{\omega_x} \;,
\label{Eq:WhyPauliLind-1Q SX+GenDiss lambda_y}\\
& \lambda_z = \frac{2\theta - \sin(2\theta)}{16}\frac{\beta_{\downarrow}}{\omega_x} + \frac{2\theta + \sin(2\theta)}{8}\frac{\beta_{\phi}}{\omega_x} \;.
\label{Eq:WhyPauliLind-1Q SX+GenDiss lambda_z}
\end{align} 
\end{subequations}

Note the dependence of $\lambda_{y}$ and $\lambda_{z}$ on $\theta$ as $2\theta \pm \sin(2\theta)$ which pre-multiply both the damping and dephasing rates. This originates from the fact that the ideal gate rotation is along the $X$ axis, which does not commute with amplitude damping or pure dephasing noise, and therefore mixes the $Y$ and $Z$ noise components.
   
\subsection{Two-qubit $CZ_{\theta}$ gate with amplitude damping and pure dephasing}
\label{SubApp:CZ_th+AD+PD}
Consider a $CZ_{\theta}$ gate with amplitude damping and pure dephasing on each qubit with the starting Lindblad equation
\begin{align}
\begin{split}
\dot{\rho}(t) & = - i \left[\frac{\omega_{cz}}{2}(II-IZ-ZI+ZZ), \rho(t)\right] \\
& + \sum\limits_{j=l,r} \beta_{\downarrow j} \mathcal{D}[S_j^{-}] \rho(t) 
+ \sum\limits_{j=l,r} \frac{\beta_{\phi j}}{2} \mathcal{D}[Z_j] \rho(t)  \;.
\end{split}
\label{Eq:WhyPauliLind-2Q CZ+GenDiss Lind}
\end{align}
The ideal evolution in the interval $[0, \tau_g]$ for $\tau_g=\theta/\omega_{cz}$ results in a partial-angle $CZ_{\theta}\equiv \exp[-i(\theta/2)(II-IZ-ZI+ZZ)]$ gate. 

Following the same steps as in Sec.~\ref{SubApp:X_th+AD+PD}, up to the leading order in $\beta_{\downarrow j}/\omega_{cz}$ and $\beta_{\phi j}/\omega_{cz}$, for $j=l,r$, the twirled noise channel can be generated via a PL model with the following non-zero model coefficients: 
\begin{subequations}
\begin{align}
& \lambda_{iz} = \frac{\theta}{2} \frac{\beta_{\phi r}}{\omega_{cz}} \;, 
\label{Eq:WhyPauliLind-2Q CZ+GenDiss-lamb_iz}\\
& \lambda_{zi} = \frac{\theta}{2} \frac{\beta_{\phi l}}{\omega_{cz}}  \;, 
\label{Eq:WhyPauliLind-2Q CZ+GenDiss-lamb_zi}\\
& \lambda_{ix} = \lambda_{iy} = \frac{2\theta+\sin(2\theta)}{16}\frac{\beta_{\downarrow r}}{\omega_{cz}} \;, 
\label{Eq:WhyPauliLind-2Q CZ+GenDiss-lamb_ix}\\
& \lambda_{xi} = \lambda_{yi} = \frac{2\theta+\sin(2\theta)}{16}\frac{\beta_{\downarrow l}}{\omega_{cz}} \;, 
\label{Eq:WhyPauliLind-2Q CZ+GenDiss-lamb_xi}\\
& \lambda_{zx} = \lambda_{zy} = \frac{2\theta-\sin(2\theta)}{16}\frac{\beta_{\downarrow r}}{\omega_{cz}} \;, 
\label{Eq:WhyPauliLind-2Q CZ+GenDiss-lamb_zx}\\
& \lambda_{xz} = \lambda_{yz} = \frac{2\theta-\sin(2\theta)}{16}\frac{\beta_{\downarrow l}}{\omega_{cz}} \;. 
\label{Eq:WhyPauliLind-2Q CZ+GenDiss-lamb_xz}
\end{align}
\end{subequations}
Importantly, despite the local (weight-1) form of the starting amplitude damping and dephasing noise, the twirled noise channel has weight-2 Pauli terms as in Eqs.~(\ref{Eq:WhyPauliLind-2Q CZ+GenDiss-lamb_ix})--(\ref{Eq:WhyPauliLind-2Q CZ+GenDiss-lamb_xz}). Moreover, the gate angle dependence $2\theta \pm \sin(2\theta)$ re-appears as in the $X_{\theta}$ gate example of Sec.~\ref{SubApp:X_th+AD+PD}. For an arbitrary $\theta$, there is a two-fold degeneracy of model parameters according to Eqs.~(\ref{Eq:WhyPauliLind-2Q CZ+GenDiss-lamb_ix})--(\ref{Eq:WhyPauliLind-2Q CZ+GenDiss-lamb_xz}), whereas for Clifford angles $\theta = n\pi/2$ the degeneracy becomes four-fold where $\lambda_{ix}=\lambda_{iy}=\lambda_{zx}=\lambda_{zy}$ and $\lambda_{xi}=\lambda_{yi}=\lambda_{xz}=\lambda_{yz}$. 

\subsection{Two-qubit $CX_{\theta}$ gate with amplitude damping and pure dephasing}
\label{SubApp:CX_th+AD+PD}
We next consider a $CX_{\theta}$ gate with amplitude damping and pure dephasing on each qubit with the starting Lindblad model:
\begin{align}
\begin{split}
\dot{\rho}(t) & = - i \left[\frac{\omega_{cx}}{2}(IX-ZX), \rho(t)\right] \\
& + \sum\limits_{j=l,r} \beta_{\downarrow j} \mathcal{D}[S_j^{-}] \rho(t) 
+ \sum\limits_{j=l,r} \frac{\beta_{\phi j}}{2} \mathcal{D}[Z_j] \rho(t)  \;.
\end{split}
\label{Eq:WhyPauliLind-2Q Cx+GenDiss Lind}
\end{align}
The ideal evolution for $\tau_g=\theta/\omega_{cx}$ gives a partial-angle $CX_{\theta}\equiv \exp[-i(\theta/2)(IX-ZX)]$ gate, which is locally equivalent to a CNOT gate at $\theta = \pi/2$.

Following the same steps as in Sec.~\ref{SubApp:CZ_th+AD+PD}, we compute the PL model parameters up to the lowest order in $\beta_{\downarrow j}$ and $\beta_{\phi j}$ for $j=l,r$ as:
\begin{subequations}
\begin{align}
\lambda_{ix} & = \frac{\theta}{4}\frac{\beta_{\downarrow r}}{\omega_{cx}} \;,
\label{Eq:WhyPauliLind-2Q Cx+GenDiss-lamb_ix}\\
\begin{split}
\lambda_{iy} & = \frac{12\theta + 8\sin(2\theta)+\sin(4\theta)}{128}\frac{\beta_{\downarrow r}}{\omega_{cx}} \\
& + \frac{4\theta - \sin(4\theta)}{64}\frac{\beta_{\phi r}}{\omega_{cx}} \;,
\end{split} \\
\begin{split}
\lambda_{iz} & = \frac{4\theta-\sin(4\theta)}{128}\frac{\beta_{\downarrow r}}{\omega_{cx}} \\
& + \frac{12\theta + 8\sin(2\theta)+\sin(4\theta)}{64}\frac{\beta_{\phi r}}{\omega_{cx}}\;,
\end{split}\\
\lambda_{xi} & = \lambda_{yi} = \frac{2\theta+\sin(2\theta)}{16}\frac{\beta_{\downarrow l}}{\omega_{cx}} \;,\\
\lambda_{xx} & = \lambda_{yx} = \frac{2\theta-\sin(2\theta)}{16}\frac{\beta_{\downarrow l}}{\omega_{cx}} \;,\\
\lambda_{zi} & = \frac{\theta}{2} \frac{\beta_{\phi l}}{\omega_{cx}} \;,\\
\begin{split}
\lambda_{zy} & = \frac{12\theta - 8\sin(2\theta)+\sin(4\theta)}{128}\frac{\beta_{\downarrow r}}{\omega_{cx}} \\
& + \frac{4\theta - \sin(4\theta)}{64}\frac{\beta_{\phi r}}{\omega_{cx}}\;,
\end{split} \\
\begin{split}
\lambda_{zz} & = \frac{4\theta-\sin(4\theta)}{128}\frac{\beta_{\downarrow r}}{\omega_{cx}} \\
& + \frac{12\theta - 8\sin(2\theta)+\sin(4\theta)}{64}\frac{\beta_{\phi r}}{\omega_{cx}}\;,
\end{split}
\label{Eq:WhyPauliLind-2Q Cx+GenDiss-lamb_zz}
\end{align}
\end{subequations}

The results for amplitude damping and pure dephasing in Secs.~\ref{SubApp:CZ_th+AD+PD} and~\ref{SubApp:CX_th+AD+PD} are summarized in Tables~\ref{Tab:MagnusPert-T1Decay} and~\ref{Tab:MagnusPert-T2Decay}, respectively. 

\subsection{Two-qubit phase noise}
\label{SubApp:2QPhNoise}

We next study effective PL models for coherent noise under a two-qubit operation. Consider the Lindblad equation
\begin{align}
\dot{\rho}(t) = - i \left[H_g + H_{\delta}, \rho(t)\right] \;,
\label{Eq:WhyPauliLind-2Q Cx+PhaseNoise}
\end{align}
where $H_{\delta}$ is the phase noise with $IZ$, $ZI$ and $ZZ$ components as
\begin{align}
H_{\delta} \equiv \frac{\delta_{iz}}{2} IZ + \frac{\delta_{zi}}{2} ZI + \frac{\delta_{zz}}{2} ZZ \;.
\label{Eq:WhyPauliLind-2Q Cx+PhaseNoise-Def of Hdelta}
\end{align}
We consider three cases: (i) $I_{\tau_g}$ with $H_{g}=0$, (ii) $CZ_{\theta}$ with $H_{g}=(\omega_{cz}/2)(II-IZ-ZI+ZZ)$, and finally (iii) $CX_{\theta}$ with $H_{g}=(\omega_{cz}/2)(IX-ZX)$.

We find that the leading contribution to the PL noise model from coherent noise is second order in $\delta_{k}$, for $k=iz$, $zi$ and $zz$. In particular, among the second-order terms in the Dyson series~(\ref{Eq:TDLindPT-N_Ch Dyson Sol}) for the noise channel, the $\Omega_2(\tau_g,0)$ term contributes only to off-diagonal (non-Pauli) channel terms, which cancels out when twirled, while the $\Omega_1^2(\tau_g,0)/2$ term yields effective non-zero PL terms.

For the $I_{\tau_g}$ and $CZ_{\theta}$ gates, the noise and ideal Hamiltonians commute. The leading-order results is therefore trivial, as each Hamiltonian component would only appear in the corresponding PL parameter as 
\begin{align}
\lambda_{iz} = \frac{\theta_{cz}^2}{4}\frac{\delta_{iz}^2}{\omega_{cz}^2}, \; \lambda_{zi} = \frac{\theta_{cz}^2}{4}\frac{\delta_{zi}^2}{\omega_{cz}^2}, \; \lambda_{zz} = \frac{\theta_{cz}^2}{4}\frac{\delta_{zz}^2}{\omega_{cz}^2} \;.
\end{align}

For the case of the $CX_{\theta}$ gate, the phase noise does \textit{not} commute with the $X$ operation on the target, giving rise to a mixing of PL parameters as
\begin{align}
&\lambda_{zi} = \frac{\theta_{cz}^2}{4}\frac{\delta_{zi}^2}{\omega_{cz}^2} \;, \\
& \lambda_{iy} = \lambda_{zy}=\frac{\sin^4(\theta_{g})}{16}\frac{(\delta_{iz}-\delta_{zz})^2}{\omega_{cx}^2} \;, \\
& \lambda_{iz} = \frac{[2\theta (\delta_{iz}+\delta_{zz}) +  \sin(2\theta_{g})(\delta_{iz}-\delta_{zz})]^2}{64 \omega_{cx}^2} \;, \\
& \lambda_{zz} = \frac{[2\theta (\delta_{iz}+\delta_{zz}) + \sin(2\theta_{g}) (\delta_{zz}-\delta_{iz})]^2}{64 \omega_{cx}^2} \;.
\end{align}
The result of this section is summarized in Table~\ref{Tab:MagnusPert-2QPhaseError}.

\begin{figure*}[t!]
\centering
\includegraphics[scale=0.35]{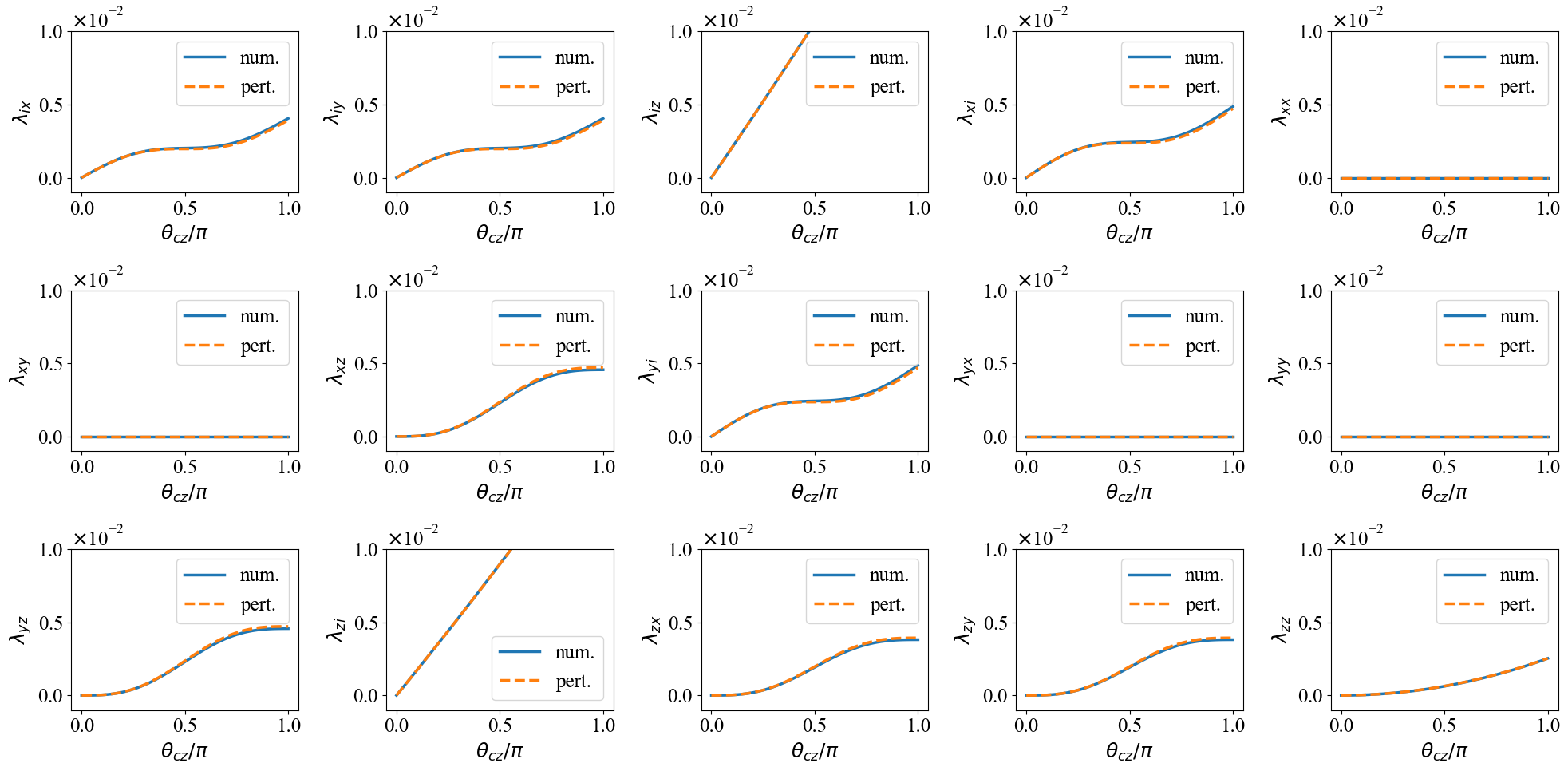}
\caption{\textbf{Dependence of PL noise generator $\lambda_k$ parameters on the angle of $CZ_{\theta}$ operation} -- The panels show the 15 PL $\lambda_k$ parameters for the two-qubit circuit in Fig.~\ref{fig:PhysNoise-2QCompToNum}(a) with coherent phase noise, amplitude damping and pure dephasing. The solid blue curve shows the numerically derived values obtained by solving the Lindblad equation with all noise sources acting simultaneously. The dashed orange curves show the sum over perturbative values in Tables~\ref{Tab:MagnusPert-T1Decay}--\ref{Tab:MagnusPert-2QPhaseError}. Gate angle $\theta$ is swept over the interval $[0,\pi]$. Other system parameters are the same as in Fig.~\ref{fig:PhysNoise-2QCompToNum}.}
\label{Fig:cz_theta_PL_angle_dependence.png}
\end{figure*}

\subsection{Three-qubit $ZZ$ crosstalk}
\label{SubApp:3QXtalk}

We next consider crosstalk in a three-qubit scenario, with a coherent $ZZ$ interaction between a third spectator qubit and either the control or the target of the gate. We again analyze three cases: (i) $I_{\tau_g}$, (ii) $CZ_{\theta}$, and finally (iii) $CX_{\theta}$ with the spectator coupled to either the control or the target.  These scenarios are summarized in Table~\ref{Tab:MagnusPert-3QZZCrosstalk}.

The only non-trivial case is that of the $CX_{\theta}\otimes I$ operation with $IZZ$ crosstalk between the target and a spectator qubit. The ideal and noise Hamiltonians in this case are $H_{g}=(\omega_{cz}/2)(IXI-ZXI)$ and $H_{\delta} = (\delta_{izz}/2)IZZ$. This leads to effective weight-2 and weight-3 PL parameters 
\begin{align}
& \lambda_{iyz}=\lambda_{zyz}=\frac{\sin^4(\theta)}{16} \frac{\delta_{izz}^2}{\omega_{cx}^2} \;, \\
& \lambda_{izz} = \frac{[2\theta+\sin(2\theta)]^2}{64} \frac{\delta_{izz}^2}{\omega_{cx}^2} \;, \\
& \lambda_{zzz} = \frac{[2\theta-\sin(2\theta)]^2}{64} \frac{\delta_{izz}^2}{\omega_{cx}^2} \;.
\end{align}
We note that the standard sparse PL implementation assumes only up to weight-2 terms, and this case serves as the simplest physical counterexample of the weight-2 sparsity assumptions. Our method can generalize the sparsity requirements by narrowing down what higher-weight PL terms can emerge.

\begin{figure*}[t!]
\centering
\includegraphics[scale=0.35]{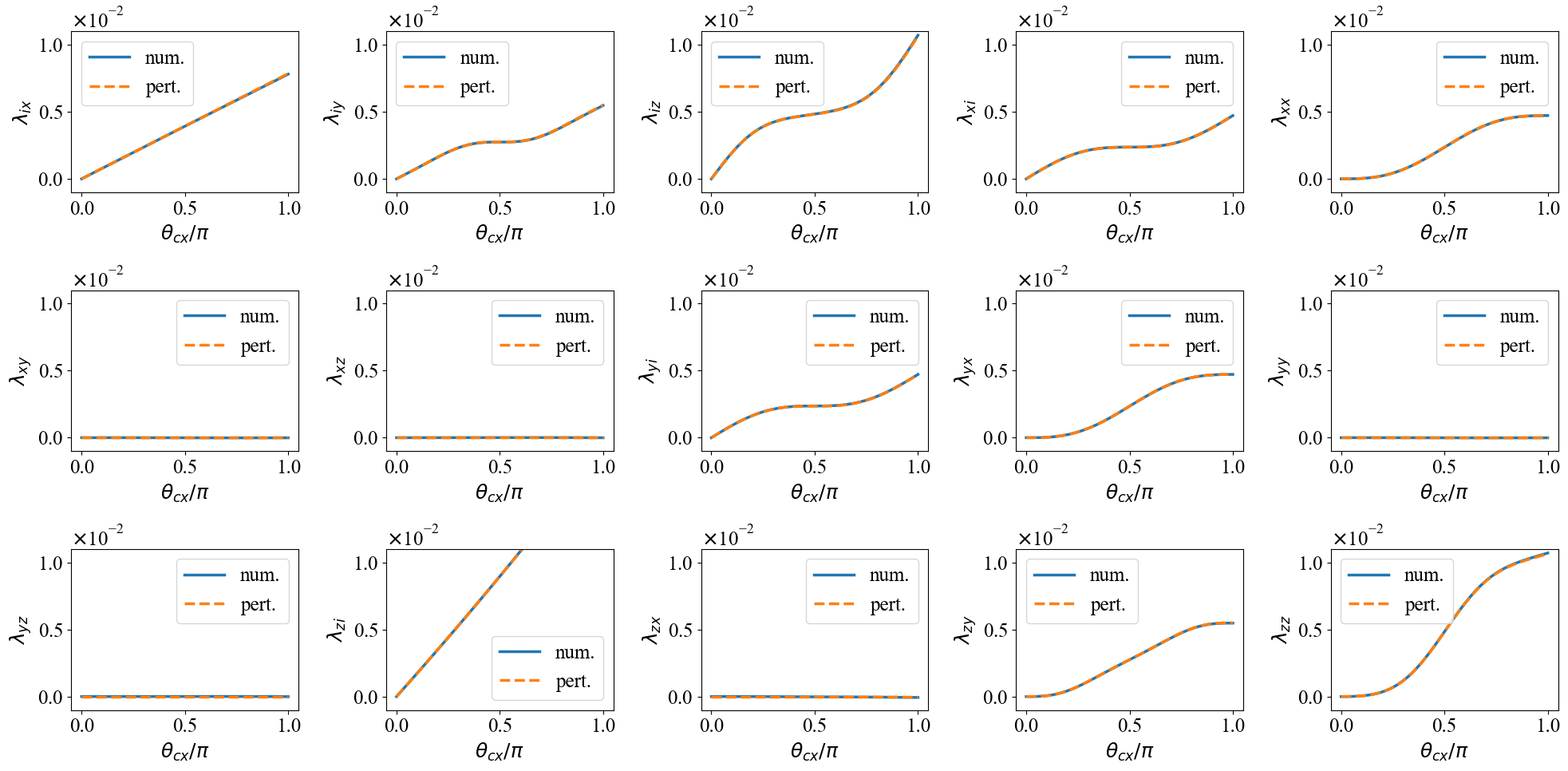}
\caption{\textbf{Dependence of PL noise generator $\lambda_k$ parameters on the angle of $CX_{\theta}$ operation} -- The panels show the 15 PL $\lambda_k$ parameters for the two-qubit circuit in Fig.~\ref{fig:PhysNoise-2QCompToNum}(f) with coherent phase noise, amplitude damping and pure dephasing. The solid blue curve shows the numerically derived values obtained by solving the Lindblad equation with all noise sources acting simultaneously. The dashed orange curves show the sum over perturbative values in Tables~\ref{Tab:MagnusPert-T1Decay}--\ref{Tab:MagnusPert-2QPhaseError}. Gate angle $\theta$ is swept in $[0,\pi]$. Other system parameters are the same as in Fig.~\ref{fig:PhysNoise-2QCompToNum}.}
\label{Fig:cx_theta_PL_angle_dependence}
\end{figure*}

\begin{figure*}[t!]
\centering
\includegraphics[scale=0.35]{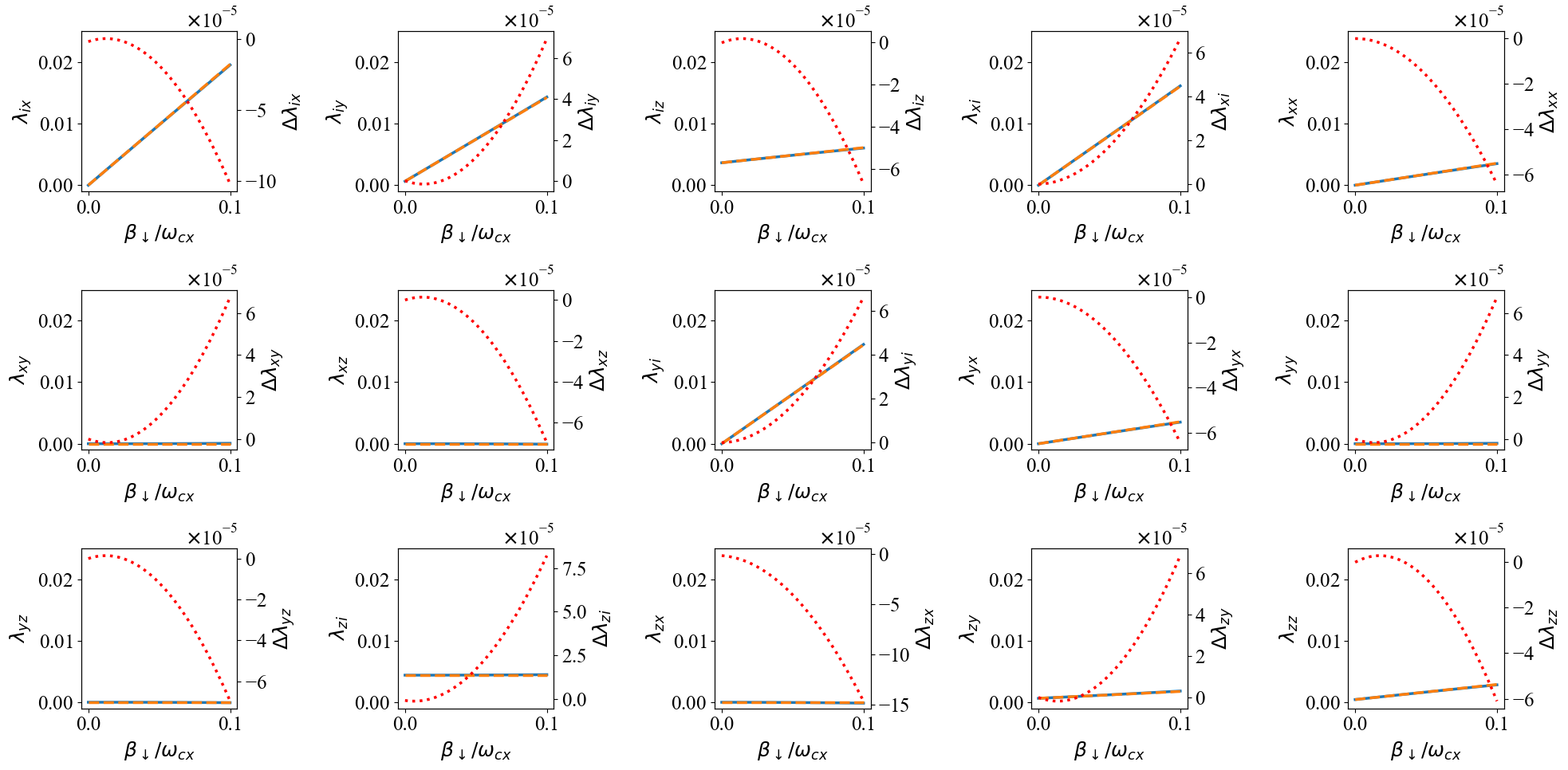}
\caption{\textbf{Precision of the leading-order perturbative PL noise model} -- For the case of a $CX_{\pi/4}$ gate as in Fig.~\ref{fig:PhysNoise-2QCompToNum}(f), the panels show the a comparison of effective PL generator parameters between numerical (solid blue) and leading-order perturbative results (dashed orange), found by the sum of Tables~\ref{Tab:MagnusPert-T1Decay}--\ref{Tab:MagnusPert-2QPhaseError}, as a function of amplitude damping relative noise strength $\beta_{\downarrow}/\omega_{cx}$ in the range $[0,0.1]$. Other system parameters are the same as in Fig.~\ref{fig:PhysNoise-2QCompToNum}. The left y axis shows the parameter values, while the right y axis and the dotted red lines show the deviation between leading-order perturbation and numerics.}
\label{Fig:first_order_PL_validity_cx_theta}
\end{figure*}
 
\subsection{Four-qubit ZZ crosstalk}
\label{SubApp:4QXtalk}

We finally analyze crosstalk in four-qubit scenarios with two adjacent two-qubit gates impacted by $ZZ$ crosstalk on the middle two qubits. We analyze four cases: (i) $I_{\tau_g}$, (ii) $CZ_{\theta}\otimes CZ_{\theta}$, (iii) $CX_{\theta}\otimes CX_{\theta}$, and (iv) finally $CX_{\theta}\otimes X_{\theta}C$. The difference between cases (iii) and (iv) is in the relative direction of the $CX_{\theta}$ gate as depicted in Fig.~\ref{fig:PhysNoise-ZZCrosstalkCases}.

Cases (iii) and (iv) yield non-trivial PL $\lambda_k$ parameters since the Hamiltonian noise term $IZZI$ does not commute with $X_{\theta}$ on target qubits. Case (iii) is effectively a three-qubit crosstalk problem, which is the same as the $CX_{\theta}\otimes I$ example of Appendix~\ref{SubApp:3QXtalk}. Case (iv), due to the crosstalk between two adjacent targets, can lead to more involved weight-3 and weight-4 PL parameters. In particular, the ideal and noise Hamiltonian for case (iv) can be written as
\begin{align}
H_g &= \frac{\omega_{cx}}{2}[(IXII-ZXII)+ (IIIX-IIZX)] \;, \\
H_{\delta}  &= \frac{\delta_{izzi}}{2}IZZI \;,
\end{align}  
resulting in the following PL parameters:
\begin{align}
\begin{split}
\lambda_{iyyi} &= \lambda_{iyyz} = \lambda_{iyyz}=\lambda_{izzz} = \lambda_{zyyi} = \lambda_{zyyz} \\
&= \frac{[4\theta - \sin(4\theta)]^2}{4096}\frac{\delta_{izzi}^2}{\omega_{cx}^2} \;,
\end{split}\\
\begin{split}
\lambda_{iyzi} &= \lambda_{izyi} = \lambda_{iyzz} = \lambda_{zzyi} \\
&= \frac{\sin^4(\theta)[3+\cos(2\theta)]^2}{256}\frac{\delta_{izzi}^2}{\omega_{cx}^2} \;,
\end{split}
\end{align}
\begin{align}
\begin{split}
\lambda_{iyzz} = \lambda_{zyzz} = \lambda_{zzyi} = \lambda_{zzyz} = \frac{\sin^8(\theta)}{64}\frac{\delta_{izzi}^2}{\omega_{cx}^2} \;,
\end{split}
\end{align}
\begin{align}
\lambda_{izzi} = \frac{[12\theta+8\sin(2\theta)+\sin(4\theta)]^2}{4096}\frac{\delta_{izzi}^2}{\omega_{cx}^2} \;,\\
\lambda_{zzzz} = \frac{[12\theta-8\sin(2\theta)+\sin(4\theta)]^2}{4096}\frac{\delta_{izzi}^2}{\omega_{cx}^2}\;.
\end{align}

\section{Dependence of PL parameters on the gate angle}
\label{App:AngleDepofPLGen}

We explore the dependence of the PL noise model parameters on the gate angle $\theta$ more thoroughly. Tables~\ref{Tab:MagnusPert-T1Decay}--\ref{Tab:MagnusPert-4QZZCrosstalk} provide the leading-order expressions for such dependence. In Figs.~\ref{fig:PhysNoise-2QCompToNum} and~\ref{fig:PhysNoise-XtalkCompToNum}, we presented good agreement between perturbation and numerical simulations for $\theta=\pi/4$ corresponding to $\sqrt{CX}$ and $\sqrt{CZ}$ gates. Here, we first show further comparison to numerical simulations at arbitrary angles. Furthermore, we discuss how we can infer useful information about the nature of the physical noise from the angle dependence of the PL noise parameters. 

Figures~\ref{Fig:cz_theta_PL_angle_dependence.png} and~\ref{Fig:cx_theta_PL_angle_dependence} show the dependence of the PL $\lambda_k$ parameters on the gate angle for the two-qubit scenarios studied in Fig.~\ref{fig:PhysNoise-2QCompToNum}. We find that the leading-order perturbative expressions in Tables~\ref{Tab:MagnusPert-T1Decay}--\ref{Tab:MagnusPert-2QPhaseError} describe the angular dependence very precisely. This is in part due to our noise construction in which the interaction-frame transformation is performed exactly, while perturbation is performed in orders of \textit{relative} noise strength.

The $\theta$ dependence of PL noise parameters can be very informative about the nature of the underlying physical noise mechanism. Generally, the effective generator due to coherent noise has a leading-order quadratic dependence on angle $\theta$ (gate time $\tau_g$), whereas for incoherent noise the dependence is linear. Depending on the commutativity with the ideal operation, the weight of the physical noise can change and transform onto other Pauli terms, which appear as a sinusoidal dependence on $\theta$. For instance, a linear $\theta$ dependence for $\lambda_{iz}$ and $\lambda_{zi}$ is a potential signature of a dominant dephasing error that has \textit{not} spilled over other terms since it commutes with the $CZ_{\theta}$ operation (Fig.~\ref{Fig:cz_theta_PL_angle_dependence.png}). However, under a $CX_{\theta}$ operation, $\lambda_{zi}$ is linear while $\lambda_{iz}$ has sinusoidal oscillations on top of a linear dependence (Fig.~\ref{Fig:cz_theta_PL_angle_dependence.png}). On the other hand, unlike the $CZ_{\theta}$ case, for $CX_{\theta}$, one finds linear $\lambda_{ix}$ as the $T_1$ noise on the target qubit commutes with the ideal operation. Furthermore, we observe a quadratic (linear+sinusoidal) dependence of the $\lambda_{zz}$ parameter for the $CZ_{\theta}$ ($CX_{\theta}$) operations, indicating the dominant underlying physical mechanism to be coherent $ZZ$ (incoherent $T_1$ and $T_{2\phi}$) noise. 

\section{Precision of leading-order Lindblad perturbation}
\label{App:LindPertPrecision}

For brevity, in Appendix~\ref{App:WhyPauliLind}, we have presented only the leading-order perturbative expressions. Our comparisons with exact numerical PL models in Figs.~\ref{fig:PhysNoise-2QCompToNum} and~\ref{fig:PhysNoise-XtalkCompToNum} shows a surprisingly good agreement with such leading-order models. Here, we quantify the deviation of the leading-order analytical results as a function of noise strength. 

For the two-qubit example of $CX_{\pi/4}$, examined in the bottom row of Fig.~\ref{fig:PhysNoise-2QCompToNum}, Fig.~\ref{Fig:first_order_PL_validity_cx_theta} shows a sweep of the amplitude damping relative noise strength $\beta_{\downarrow}/\omega_{cx}$ in the range $[0,0.1]$ while keeping other parameters the same. We find that the deviation between the leading-order perturbative expressions and the numerical values is about $O(10^{-5})$ for relative noise strength of $\beta_{\downarrow}/\omega_{cx}\approx O(10^{-2})$, but this can grow up to $O(10^{-4})$ for $\beta_{\downarrow}/\omega_{cx}\approx O(10^{-1})$.       

\bibliographystyle{unsrt}
\bibliography{LindbladNoiseAnalysis.bib}

\begin{thebibliography}{10}

\bibitem{Temme_Error_2017}
Kristan Temme, Sergey Bravyi, and Jay~M Gambetta.
\newblock Error mitigation for short-depth quantum circuits.
\newblock {\em Physical review letters}, 119(18):180509, 2017.

\bibitem{Li_Efficient_2017}
Ying Li and Simon~C Benjamin.
\newblock Efficient variational quantum simulator incorporating active error
  minimization.
\newblock {\em Physical Review X}, 7(2):021050, 2017.

\bibitem{Endo_practical_2018}
Suguru Endo, Simon~C Benjamin, and Ying Li.
\newblock Practical quantum error mitigation for near-future applications.
\newblock {\em Physical Review X}, 8(3):031027, 2018.

\bibitem{Cai_Quantum_2023}
Zhenyu Cai, Ryan Babbush, Simon~C Benjamin, Suguru Endo, William~J Huggins,
  Ying Li, Jarrod~R McClean, and Thomas~E O’Brien.
\newblock Quantum error mitigation.
\newblock {\em Reviews of Modern Physics}, 95(4):045005, 2023.

\bibitem{Kandala_Error_2019}
Abhinav Kandala, Kristan Temme, Antonio~D C{\'o}rcoles, Antonio Mezzacapo,
  Jerry~M Chow, and Jay~M Gambetta.
\newblock Error mitigation extends the computational reach of a noisy quantum
  processor.
\newblock {\em Nature}, 567(7749):491--495, 2019.

\bibitem{Kim_Scalable_2023}
Youngseok Kim, Christopher~J Wood, Theodore~J Yoder, Seth~T Merkel, Jay~M
  Gambetta, Kristan Temme, and Abhinav Kandala.
\newblock Scalable error mitigation for noisy quantum circuits produces
  competitive expectation values.
\newblock {\em Nature Physics}, 19(5):752--759, 2023.

\bibitem{Berg_Probabilistic_2023}
Ewout van~den Berg, Zlatko~K Minev, Abhinav Kandala, and Kristan Temme.
\newblock Probabilistic error cancellation with sparse {P}auli--{L}indblad
  models on noisy quantum processors.
\newblock {\em Nature Physics}, pages 1--6, 2023.

\bibitem{Kim_Evidence_2023}
Youngseok Kim, Andrew Eddins, Sajant Anand, Ken~Xuan Wei, Ewout van Den~Berg,
  Sami Rosenblatt, Hasan Nayfeh, Yantao Wu, Michael Zaletel, Kristan Temme,
  et~al.
\newblock Evidence for the utility of quantum computing before fault tolerance.
\newblock {\em Nature}, 618(7965):500--505, 2023.

\bibitem{Gupta_Probabilistic_2023}
Riddhi~S Gupta, Ewout van~den Berg, Maika Takita, Kristan Temme, and Abhinav
  Kandala.
\newblock Probabilistic error cancellation for measurement-based quantum
  circuits.
\newblock {\em arXiv preprint arXiv:2310.07825}, 2023.

\bibitem{Filippov_Scalable_2023}
Sergei Filippov, Matea Leahy, Matteo A.~C. Rossi, and Guillermo
  Garc{\'\i}a-P{\'e}rez.
\newblock Scalable tensor-network error mitigation for near-term quantum
  computing.
\newblock {\em arXiv preprint arXiv:2307.11740}, 2023.

\bibitem{Fischer_Dynamical_2024}
Laurin~E Fischer, Matea Leahy, Andrew Eddins, Nathan Keenan, Davide Ferracin,
  Matteo A.~C. Rossi, Youngseok Kim, Andre He, Francesca Pietracaprina, Boris
  Sokolov, et~al.
\newblock Dynamical simulations of many-body quantum chaos on a quantum
  computer.
\newblock {\em arXiv preprint arXiv:2411.00765}, 2024.

\bibitem{Leung_Approximate_1997}
Debbie~W Leung, Michael~A Nielsen, Isaac~L Chuang, and Yoshihisa Yamamoto.
\newblock Approximate quantum error correction can lead to better codes.
\newblock {\em Physical Review A}, 56(4):2567, 1997.

\bibitem{Fletcher_Optimum_2007}
Andrew~S Fletcher, Peter~W Shor, and Moe~Z Win.
\newblock Optimum quantum error recovery using semidefinite programming.
\newblock {\em Physical Review A—Atomic, Molecular, and Optical Physics},
  75(1):012338, 2007.

\bibitem{Nickerson_Analysing_2019}
Naomi~H Nickerson and Benjamin~J Brown.
\newblock Analysing correlated noise on the surface code using adaptive
  decoding algorithms.
\newblock {\em Quantum}, 3:131, 2019.

\bibitem{Schwartzman_Modeling_2024}
Zohar Schwartzman-Nowik, Liran Shirizly, and Haggai Landa.
\newblock Modeling error correction with {L}indblad dynamics and approximate
  channels.
\newblock {\em arXiv preprint arXiv:2402.16727}, 2024.

\bibitem{Tuckett_Tailoring_2019}
David~K Tuckett, Andrew~S Darmawan, Christopher~T Chubb, Sergey Bravyi,
  Stephen~D Bartlett, and Steven~T Flammia.
\newblock Tailoring surface codes for highly biased noise.
\newblock {\em Physical Review X}, 9(4):041031, 2019.

\bibitem{Puri_Bias_2020}
Shruti Puri, Lucas St-Jean, Jonathan~A Gross, Alexander Grimm, Nicholas~E
  Frattini, Pavithran~S Iyer, Anirudh Krishna, Steven Touzard, Liang Jiang,
  Alexandre Blais, et~al.
\newblock Bias-preserving gates with stabilized cat qubits.
\newblock {\em Science advances}, 6(34):eaay5901, 2020.

\bibitem{Sahay_High_2023}
Kaavya Sahay, Junlan Jin, Jahan Claes, Jeff~D Thompson, and Shruti Puri.
\newblock High-threshold codes for neutral-atom qubits with biased erasure
  errors.
\newblock {\em Physical Review X}, 13(4):041013, 2023.

\bibitem{Poyatos_Complete_1997}
JF~Poyatos, J~Ignacio Cirac, and Peter Zoller.
\newblock Complete characterization of a quantum process: the two-bit quantum
  gate.
\newblock {\em Physical Review Letters}, 78(2):390, 1997.

\bibitem{Chuang_Prescription_1997}
Isaac~L Chuang and Michael~A Nielsen.
\newblock Prescription for experimental determination of the dynamics of a
  quantum black box.
\newblock {\em Journal of Modern Optics}, 44(11-12):2455--2467, 1997.

\bibitem{DAriano_Quantum_2001}
G.~M. D'Ariano and P.~Lo~Presti.
\newblock Quantum tomography for measuring experimentally the matrix elements
  of an arbitrary quantum operation.
\newblock {\em Physical review letters}, 86(19):4195, 2001.

\bibitem{Mohseni_Quantum_2008}
Masoud Mohseni, Ali~T Rezakhani, and Daniel~A Lidar.
\newblock Quantum-process tomography: Resource analysis of different
  strategies.
\newblock {\em Physical Review A—Atomic, Molecular, and Optical Physics},
  77(3):032322, 2008.

\bibitem{Merkel_Self_2013}
Seth~T Merkel, Jay~M Gambetta, John~A Smolin, Stefano Poletto, Antonio~D
  C{\'o}rcoles, Blake~R Johnson, Colm~A Ryan, and Matthias Steffen.
\newblock Self-consistent quantum process tomography.
\newblock {\em Physical Review A—Atomic, Molecular, and Optical Physics},
  87(6):062119, 2013.

\bibitem{Greenbaum_Introduction_2015}
Daniel Greenbaum.
\newblock Introduction to quantum gate set tomography.
\newblock {\em arXiv preprint arXiv:1509.02921}, 2015.

\bibitem{Bennett_Purification_1996}
Charles~H. Bennett, Gilles Brassard, Sandu Popescu, Benjamin Schumacher,
  John~A. Smolin, and William~K. Wootters.
\newblock Purification of noisy entanglement and faithful teleportation via
  noisy channels.
\newblock {\em Phys. Rev. Lett.}, 76:722--725, Jan 1996.

\bibitem{Knill_Fault_2004}
Emanuel Knill.
\newblock Fault-tolerant postselected quantum computation: Threshold analysis.
\newblock arXiv:quant-ph/0404104, 2004.

\bibitem{Kern_Quantum_2005}
Oliver Kern, Gernot Alber, and Dima~L Shepelyansky.
\newblock Quantum error correction of coherent errors by randomization.
\newblock {\em The European Physical Journal D-Atomic, Molecular, Optical and
  Plasma Physics}, 32(1):153--156, 2005.

\bibitem{Geller_Efficient_2013}
Michael~R Geller and Zhongyuan Zhou.
\newblock Efficient error models for fault-tolerant architectures and the
  {P}auli twirling approximation.
\newblock {\em Physical Review A}, 88(1):012314, 2013.

\bibitem{Wallman_Noise_2016}
Joel~J Wallman and Joseph Emerson.
\newblock Noise tailoring for scalable quantum computation via randomized
  compiling.
\newblock {\em Physical Review A}, 94(5):052325, 2016.

\bibitem{Samach_Lindblad_2022}
Gabriel~O Samach, Ami Greene, Johannes Borregaard, Matthias Christandl, Joseph
  Barreto, David~K Kim, Christopher~M McNally, Alexander Melville, Bethany~M
  Niedzielski, Youngkyu Sung, et~al.
\newblock Lindblad tomography of a superconducting quantum processor.
\newblock {\em Physical Review Applied}, 18(6):064056, 2022.

\bibitem{Pastori_Characterization_2022}
Lorenzo Pastori, Tobias Olsacher, Christian Kokail, and Peter Zoller.
\newblock Characterization and verification of {T}rotterized digital quantum
  simulation via {H}amiltonian and {L}iouvillian learning.
\newblock {\em PRX Quantum}, 3(3):030324, 2022.

\bibitem{Franca_Efficient_2024}
Daniel Stilck~Fran{\c{c}}a, Liubov~A Markovich, VV~Dobrovitski, Albert~H
  Werner, and Johannes Borregaard.
\newblock Efficient and robust estimation of many-qubit {H}amiltonians.
\newblock {\em Nature Communications}, 15(1):311, 2024.

\bibitem{Olsacher_Hamiltonian_2024}
Tobias Olsacher, Tristan Kraft, Christian Kokail, Barbara Kraus, and Peter
  Zoller.
\newblock Hamiltonian and {L}iouvillian learning in weakly-dissipative quantum
  many-body systems.
\newblock {\em arXiv preprint arXiv:2405.06768}, 2024.

\bibitem{IBM_fractionalgates_2024}
Daniella~Garcia Almeida, Kaelyn Ferris, Naoki Kanazawa, Blake Johnson, and
  Blake Davis.
\newblock New fractional gates reduce circuit depth for utility-scale
  workloads.
\newblock \url{https://www.ibm.com/quantum/blog/fractional-gates}, 2024.

\bibitem{Layden_Theory_2024}
David Layden, Bradley Mitchell, and Karthik Siva.
\newblock Theory of quantum error mitigation for non-{C}lifford gates.
\newblock {\em arXiv preprint arXiv:2403.18793}, 2024.

\bibitem{Lao_Software_2022}
Lingling Lao, Alexander Korotkov, Zhang Jiang, Wojciech Mruczkiewicz, Thomas~E
  O'Brien, and Dan~E Browne.
\newblock Software mitigation of coherent two-qubit gate errors.
\newblock {\em Quantum Science and Technology}, 7(2):025021, 2022.

\bibitem{Seif_Suppressing_2024}
Alireza Seif, Haoran Liao, Vinay Tripathi, Kevin Krsulich, Moein Malekakhlagh,
  Mirko Amico, Petar Jurcevic, and Ali Javadi-Abhari.
\newblock Suppressing correlated noise in quantum computers via context-aware
  compiling.
\newblock In {\em Proceedings of the 51st Annual International Symposium on
  Computer Architecture}. IEEE, 2024.

\bibitem{Haeberlen_Coherent_1968}
Ulrich Haeberlen and John~S Waugh.
\newblock Coherent averaging effects in magnetic resonance.
\newblock {\em Physical Review}, 175(2):453, 1968.

\bibitem{Mehring_Principles_2012}
Michael Mehring.
\newblock {\em ``Principles of high resolution NMR in solids''}.
\newblock Springer Science \& Business Media, 2012.

\bibitem{Magnus_Exponential_1954}
Wilhelm Magnus.
\newblock On the exponential solution of differential equations for a linear
  operator.
\newblock {\em Communications on pure and applied mathematics}, 7(4):649--673,
  1954.

\bibitem{Blanes_Magnus_2009}
Sergio Blanes, Fernando Casas, Jose-Angel Oteo, and Jos{\'e} Ros.
\newblock The {M}agnus expansion and some of its applications.
\newblock {\em Physics reports}, 470(5-6):151--238, 2009.

\bibitem{Blanes_Pedagogical_2010}
Sergio Blanes, Fernando Casas, Jose-Angel Oteo, and Javier Ros.
\newblock A pedagogical approach to the {M}agnus expansion.
\newblock {\em European Journal of Physics}, 31(4):907, 2010.

\bibitem{Puzzuoli_Algorithms_2023}
Daniel Puzzuoli, Sophia~Fuhui Lin, Moein Malekakhlagh, Emily Pritchett,
  Benjamin Rosand, and Christopher~J Wood.
\newblock Algorithms for perturbative analysis and simulation of quantum
  dynamics.
\newblock {\em Journal of Computational Physics}, 489:112262, 2023.

\bibitem{Dyson_SMatrixQED_1949}
Freeman~J Dyson.
\newblock The {S} matrix in quantum electrodynamics.
\newblock {\em Physical Review}, 75(11):1736, 1949.

\bibitem{Dyson_Radiation_1949}
Freeman~J Dyson.
\newblock The radiation theories of {T}omonaga, {S}chwinger, and {F}eynman.
\newblock {\em Physical Review}, 75(3):486, 1949.

\bibitem{Shillito_Fast_2021}
Ross Shillito, Jonathan~A Gross, Agustin Di~Paolo, {\'E}lie Genois, and
  Alexandre Blais.
\newblock Fast and differentiable simulation of driven quantum systems.
\newblock {\em Physical Review Research}, 3(3):033266, 2021.

\bibitem{Kim_Error_2024}
Youngseok Kim, Luke~CG Govia, Andrew Dane, Ewout van~den Berg, David~M Zajac,
  Bradley Mitchell, Yinyu Liu, Karthik Balakrishnan, George Keefe, Adam
  Stabile, et~al.
\newblock Error mitigation with stabilized noise in superconducting quantum
  processors.
\newblock {\em arXiv preprint arXiv:2407.02467}, 2024.

\bibitem{Tripathi_Deterministic_2024}
Vinay Tripathi, Daria Kowsari, Kumar Saurav, Haimeng Zhang, Eli~M
  Levenson-Falk, and Daniel~A Lidar.
\newblock Deterministic benchmarking of quantum gates.
\newblock {\em arXiv preprint arXiv:2407.09942}, 2024.

\bibitem{Blume_Taxonomy_2022}
Robin Blume-Kohout, Marcus~P da~Silva, Erik Nielsen, Timothy Proctor, Kenneth
  Rudinger, Mohan Sarovar, and Kevin Young.
\newblock A taxonomy of small {M}arkovian errors.
\newblock {\em PRX Quantum}, 3(2):020335, 2022.

\bibitem{Pedersen_Fidelity_2007}
Line~Hjortsh{\o}j Pedersen, Niels~Martin M{\o}ller, and Klaus M{\o}lmer.
\newblock Fidelity of quantum operations.
\newblock {\em Physics Letters A}, 367(1-2):47--51, 2007.

\bibitem{Abad_Universal_2022}
Tahereh Abad, Jorge Fern{\'a}ndez-Pend{\'a}s, Anton Frisk~Kockum, and G{\"o}ran
  Johansson.
\newblock Universal fidelity reduction of quantum operations from weak
  dissipation.
\newblock {\em Physical Review Letters}, 129(15):150504, 2022.

\bibitem{Gorini_completely_1976}
Vittorio Gorini, Andrzej Kossakowski, and Ennackal Chandy~George Sudarshan.
\newblock Completely positive dynamical semigroups of {$N$}-level systems.
\newblock {\em Journal of Mathematical Physics}, 17(5):821--825, 1976.

\bibitem{Lindblad_Generators_1976}
Goran Lindblad.
\newblock On the generators of quantum dynamical semigroups.
\newblock {\em Communications in mathematical physics}, 48:119--130, 1976.

\bibitem{Breuer_Theory_2002}
Heinz-Peter Breuer and Francesco Petruccione.
\newblock {\em The theory of open quantum systems}.
\newblock Oxford University Press, USA, 2002.

\bibitem{Malekakhlagh_Time_2022}
Moein Malekakhlagh, Easwar Magesan, and Luke~CG Govia.
\newblock ``time-dependent schrieffer-wolff-lindblad perturbation theory:
  Measurement-induced dephasing and second-order stark shift in dispersive
  readout''.
\newblock {\em Physical Review A}, 106(5):052601, 2022.

\bibitem{Dai_Floquet_2016}
CM~Dai, ZC~Shi, and XX~Yi.
\newblock Floquet theorem with open systems and its applications.
\newblock {\em Physical Review A}, 93(3):032121, 2016.

\bibitem{Schnell_High_2021}
Alexander Schnell, Sergey Denisov, and Andr{\'e} Eckardt.
\newblock High-frequency expansions for time-periodic {L}indblad generators.
\newblock {\em Physical Review B}, 104(16):165414, 2021.

\bibitem{Mizuta_Breakdown_2021}
Kaoru Mizuta, Kazuaki Takasan, and Norio Kawakami.
\newblock Breakdown of {M}arkovianity by interactions in stroboscopic
  {F}loquet-{L}indblad dynamics under high-frequency drive.
\newblock {\em Physical Review A}, 103(2):L020202, 2021.

\bibitem{Haddadfarshi_Completely_2015}
Farhang Haddadfarshi, Jian Cui, and Florian Mintert.
\newblock Completely positive approximate solutions of driven open quantum
  systems.
\newblock {\em Physical Review Letters}, 114(13):130402, 2015.

\bibitem{Burum_Magnus_1981}
Douglas~P Burum.
\newblock Magnus expansion generator.
\newblock {\em Physical Review B}, 24(7):3684, 1981.

\bibitem{Salzman_Alternative_1985}
WR~Salzman.
\newblock An alternative to the {M}agnus expansion in time-dependent
  perturbation theory.
\newblock {\em The Journal of chemical physics}, 82(2):822--826, 1985.

\bibitem{Paraoanu_Microwave_2006}
GS~Paraoanu.
\newblock Microwave-induced coupling of superconducting qubits.
\newblock {\em Physical Review B—Condensed Matter and Materials Physics},
  74(14):140504, 2006.

\bibitem{Rigetti_Fully_2010}
Chad Rigetti and Michel Devoret.
\newblock Fully microwave-tunable universal gates in superconducting qubits
  with linear couplings and fixed transition frequencies.
\newblock {\em Physical Review B—Condensed Matter and Materials Physics},
  81(13):134507, 2010.

\bibitem{Sheldon_Procedure_2016}
Sarah Sheldon, Easwar Magesan, Jerry~M. Chow, and Jay~M. Gambetta.
\newblock Procedure for systematically tuning up cross-talk in the
  cross-resonance gate.
\newblock {\em Phys. Rev. A}, 93:060302, Jun 2016.

\bibitem{Magesan_Effective_2020}
Easwar Magesan and Jay~M. Gambetta.
\newblock Effective {H}amiltonian models of the cross-resonance gate.
\newblock {\em Phys. Rev. A}, 101:052308, May 2020.

\bibitem{Tripathi_Operation_2019}
Vinay Tripathi, Mostafa Khezri, and Alexander~N. Korotkov.
\newblock Operation and intrinsic error budget of a two-qubit cross-resonance
  gate.
\newblock {\em Phys. Rev. A}, 100:012301, Jul 2019.

\bibitem{Malekakhlagh_First_2020}
Moein Malekakhlagh, Easwar Magesan, and David~C. McKay.
\newblock First-principles analysis of cross-resonance gate operation.
\newblock {\em Phys. Rev. A}, 102:042605, Oct 2020.

\bibitem{Kandala_Demonstration_2021}
Abhinav Kandala, Ken~X Wei, Srikanth Srinivasan, Easwar Magesan, S~Carnevale,
  GA~Keefe, D~Klaus, O~Dial, and DC~McKay.
\newblock Demonstration of a high-fidelity cnot gate for fixed-frequency
  transmons with engineered {ZZ} suppression.
\newblock {\em Physical Review Letters}, 127(13):130501, 2021.

\bibitem{Malekakhlagh_Mitigating_2022}
Moein Malekakhlagh and Easwar Magesan.
\newblock Mitigating off-resonant error in the cross-resonance gate.
\newblock {\em Physical Review A}, 105(1):012602, 2022.

\bibitem{Itoko_Three_2024}
Toshinari Itoko, Moein Malekakhlagh, Naoki Kanazawa, and Maika Takita.
\newblock Three-qubit parity gate via simultaneous cross-resonance drives.
\newblock {\em Physical Review Applied}, 21(3):034018, 2024.

\bibitem{Yan_Tunable_2018}
Fei Yan, Philip Krantz, Youngkyu Sung, Morten Kjaergaard, Daniel~L Campbell,
  Terry~P Orlando, Simon Gustavsson, and William~D Oliver.
\newblock Tunable coupling scheme for implementing high-fidelity two-qubit
  gates.
\newblock {\em Physical Review Applied}, 10(5):054062, 2018.

\bibitem{Foxen_Demonstrating_2020}
Brooks Foxen, Charles Neill, Andrew Dunsworth, Pedram Roushan, Ben Chiaro,
  Anthony Megrant, Julian Kelly, Zijun Chen, Kevin Satzinger, Rami Barends,
  et~al.
\newblock Demonstrating a continuous set of two-qubit gates for near-term
  quantum algorithms.
\newblock {\em Physical Review Letters}, 125(12):120504, 2020.

\bibitem{Collodo_Implementation_2020}
Michele~C Collodo, Johannes Herrmann, Nathan Lacroix, Christian~Kraglund
  Andersen, Ants Remm, Stefania Lazar, Jean-Claude Besse, Theo Walter, Andreas
  Wallraff, and Christopher Eichler.
\newblock Implementation of conditional phase gates based on tunable {ZZ}
  interactions.
\newblock {\em Physical review letters}, 125(24):240502, 2020.

\bibitem{Sung_Realization_2021}
Youngkyu Sung, Leon Ding, Jochen Braum{\"u}ller, Antti Veps{\"a}l{\"a}inen,
  Bharath Kannan, Morten Kjaergaard, Ami Greene, Gabriel~O Samach, Chris
  McNally, David Kim, et~al.
\newblock Realization of high-fidelity {CZ} and {ZZ}-free {iSWAP} gates with a
  tunable coupler.
\newblock {\em Physical Review X}, 11(2):021058, 2021.

\bibitem{Stehlik_Tunable_2021}
J~Stehlik, DM~Zajac, DL~Underwood, T~Phung, J~Blair, S~Carnevale, D~Klaus,
  GA~Keefe, A~Carniol, Muir Kumph, et~al.
\newblock Tunable coupling architecture for fixed-frequency transmon
  superconducting qubits.
\newblock {\em Physical review letters}, 127(8):080505, 2021.

\bibitem{Li_Realization_2024}
Rui Li, Kentaro Kubo, Yinghao Ho, Zhiguang Yan, Yasunobu Nakamura, and Hayato
  Goto.
\newblock Realization of high-fidelity {CZ} gate based on a double-transmon
  coupler.
\newblock {\em Physical Review X}, 14(4):041050, 2024.

\bibitem{Sorensen_Quantum_1999}
Anders S{\o}rensen and Klaus M{\o}lmer.
\newblock Quantum computation with ions in thermal motion.
\newblock {\em Physical review letters}, 82(9):1971, 1999.

\bibitem{Wu_noise_2018}
Yukai Wu, Sheng-Tao Wang, and L-M Duan.
\newblock Noise analysis for high-fidelity quantum entangling gates in an
  anharmonic linear paul trap.
\newblock {\em Physical Review A}, 97(6):062325, 2018.

\bibitem{Puzzuoli_Qiskit_2023}
Daniel Puzzuoli, Christopher~J Wood, Daniel~J Egger, Benjamin Rosand, and Kento
  Ueda.
\newblock Qiskit dynamics: A {P}ython package for simulating the time dynamics
  of quantum systems.
\newblock {\em Journal of Open Source Software}, 8(90):5853, 2023.

\bibitem{Chen_Learnability_2023}
Senrui Chen, Yunchao Liu, Matthew Otten, Alireza Seif, Bill Fefferman, and
  Liang Jiang.
\newblock The learnability of {P}auli noise.
\newblock {\em Nature Communications}, 14(1):52, 2023.

\bibitem{Seif_Entanglement_2024}
Alireza Seif, Senrui Chen, Swarnadeep Majumder, Haoran Liao, Derek~S Wang,
  Moein Malekakhlagh, Ali Javadi-Abhari, Liang Jiang, and Zlatko~K Minev.
\newblock Entanglement-enhanced learning of quantum processes at scale.
\newblock {\em arXiv preprint arXiv:2408.03376}, 2024.

\bibitem{Chen_Efficient_2024}
Senrui Chen, Zhihan Zhang, Liang Jiang, and Steven~T Flammia.
\newblock Efficient self-consistent learning of gate set {P}auli noise.
\newblock {\em arXiv preprint arXiv:2410.03906}, 2024.

\bibitem{Wolf_Dividing_2008}
Michael~M Wolf and J~Ignacio Cirac.
\newblock Dividing quantum channels.
\newblock {\em Communications in Mathematical Physics}, 279:147--168, 2008.

\bibitem{Wolf_Assessing_2008}
Michael~Marc Wolf, J~Eisert, Toby~S Cubitt, and J~Ignacio Cirac.
\newblock Assessing non-{M}arkovian quantum dynamics.
\newblock {\em Physical review letters}, 101(15):150402, 2008.

\bibitem{Nielsen_Guaranteed_2016}
Frank Nielsen and Ke~Sun.
\newblock Guaranteed bounds on information-theoretic measures of univariate
  mixtures using piecewise log-sum-exp inequalities.
\newblock {\em Entropy}, 18(12):442, 2016.

\bibitem{Wood_Tensor_2011}
Christopher~J Wood, Jacob~D Biamonte, and David~G Cory.
\newblock Tensor networks and graphical calculus for open quantum systems.
\newblock {\em arXiv preprint arXiv:1111.6950}, 2011.

\end{thebibliography}

\end{document}